\documentclass[showpacs,reprint,preprintnumbers,amsmath,amssymb,superscriptaddress,longbibliography,nofootinbib]{revtex4-2}

\usepackage[utf8]{inputenc}
\usepackage{amsmath}
\usepackage{amssymb} 
\usepackage{enumerate}
\usepackage{multirow}
\usepackage{braket}
\usepackage[toc,page]{appendix}
\usepackage{url}
\usepackage{graphicx}
\usepackage[pdftex,bookmarks,linktocpage,pdfpagelabels,plainpages=false,hyperfigures,linkcolor=blue,citecolor=blue]{hyperref} 
\hypersetup{colorlinks=true}
\usepackage{adjustbox}
\usepackage{mathrsfs,amssymb}  
\usepackage{cancel}
\usepackage[normalem]{ulem}
\usepackage{cleveref}
\usepackage{tikz}
\usetikzlibrary{decorations.markings}
\usetikzlibrary{positioning}
\usetikzlibrary{shapes}
\usetikzlibrary{calc}

\newcommand{\beq}{\begin{equation}}
\newcommand{\eeq}{\end{equation}}

\newcommand{\bea}{\begin{eqnarray}}
\newcommand{\eea}{\end{eqnarray}}

\newcommand\scalemath[2]{\scalebox{#1}{\mbox{\ensuremath{\displaystyle #2}}}}

\begin{document}

\title{Streamlined jet tagging network assisted by jet prong structure} 
\author{A.\,Hammad}
\email{hamed@post.kek.jp}
\affiliation{Theory Center, IPNS, KEK, 1-1 Oho, Tsukuba, Ibaraki 305-0801, Japan.}
\author{Mihoko M. \,Nojiri}
\email{nojiri@post.kek.jp }
\affiliation{Theory Center, IPNS, KEK, 1-1 Oho, Tsukuba, Ibaraki 305-0801, Japan.}
\affiliation{The Graduate University of Advanced Studies (Sokendai), 1-1 Oho, Tsukuba, Ibaraki 305-0801, Japan.}
\affiliation{Kavli IPMU (WPI), University of Tokyo, 5-1-5 Kashiwanoha, Kashiwa, Chiba 277-8583, Japan.}
\date{\today}

\begin{abstract}
Attention-based transformer models have become increasingly prevalent in collider analysis, offering enhanced performance for tasks such as jet tagging. However, they are computationally intensive and require substantial data for training. 
In this paper, we introduce a new jet classification network using an MLP mixer, where two subsequent MLP operations serve to transform particle and feature tokens over the jet constituents. The transformed particles are combined with subjet information using multi-head cross-attention so that the network is invariant under the permutation of the jet constituents. 
 We utilize two clustering algorithms to identify subjets: the standard sequential recombination algorithms with fixed radius parameters and a new IRC-safe, density-based algorithm of dynamic radii based on HDBSCAN.
The proposed network demonstrates comparable classification performance to state-of-the-art models while boosting computational efficiency drastically. Finally, we evaluate the network performance using various interpretable methods, including centred kernel alignment and attention maps, to highlight network efficacy in collider analysis tasks.
 \end{abstract}
 
\maketitle
\tableofcontents
\section{\bf Introduction}  
Machine learning techniques have exhibited significant effectiveness in various collider studies, spanning tasks such as event classification, anomaly detection and searches for new physics.  Particularly notable in high energy physics applications is identifying elementary particles that initiate jets, known as jet tagging. This is driven by the inherent differences in characteristics exhibited by jets initiated by different particles.   For instance, boosted heavy particles such as $W$, $Z$, and $H$ bosons may manifest a multi-prong structure in a jet, whereas QCD jets are less likely to exhibit such features. 
This information serves as a discriminatory factor between jets initiated by different particles. 
Jet tagging methods were initially developed based on insights of QCD theory and have undergone continuous refinement to effectively discriminate between boosted heavy objects, quark jet and gluon jet.\cite{Butterworth:2008iy,Kaplan:2008ie, Cui:2010km,Plehn:2011sj,Soper:2012pb,Anders:2013oga,Kasieczka:2015jma,Thaler:2010tr,Thaler:2011gf,Larkoski:2013eya,Moult:2016cvt,Larkoski:2014wba,Abdesselam:2010pt,Altheimer:2012mn,Altheimer:2013yza}(and references therein).
Recently,  improvements in jet identification have continued by using ML methods for jet image analysis \cite{Cogan:2014oua, Almeida:2015jua, deOliveira:2015xxd, Baldi:2016fql, Barnard:2016qma,Komiske:2016rsd,Kasieczka:2017nvn, Macaluso:2018tck,Choi:2018dag}, graph based analysis \cite{Shlomi:2020gdn,  Mokhtar:2022pwm,Ma:2022bvt,Gong:2022lye,Dreyer:2020brq} or sequence based analysis \cite{Guest:2016iqz,Pearkes:2017hku,Egan:2017ojy,Fraser:2018ieu,Butter:2017cot,Kasieczka:2018lwf}. 

Because the order of the particles in an event should not alter the outcome, any ML model for HEP should be a function of a set of particles without order, a ``particle cloud".  While these methods have demonstrated high performance in various applications, they are not well-suited for particle cloud analysis.
The particle cloud model mitigates combinatorial ambiguities by representing final state particles as a permutation-invariant sequence.
A network with an input data set of size $N$ must be invariant to $N!$ permutations of the inputs. 
This simultaneously provides the ability to capture both local structure from nearby particles and global structure resulting among all particles in the cloud by ensuring that all possible combinations of particles are considered. 
Models based on the Particle clouds were initially introduced by \cite{Komiske:2018cqr,Qu:2019gqs} in particle physics. %

These models offer not only the advantages of particle-based approaches but also the flexibility to incorporate arbitrary features of particles, such as particle ID and vertex.  

Several particle cloud ML models
have been introduced for collider analysis, including Deep Sets \cite{Komiske:2018cqr}, Edge convolution \cite{Qu:2019gqs} and Transformers \cite{Qu:2022mxj,Finke:2023veq,Shmakov:2021qdz,Hammad:2023sbd,He:2023cfc}.  
Deep Sets model is first introduced in \cite{zaheer2017deep}.  
To achieve state-of-the-art performance, the Deep Sets model requires a sizeable latent space of the network and becomes very complex. 
Edge Convolution Neural Network (EDGCNN) offers a method to incorporate local information gained 
from the nearest neighbours of each particle in the cloud for learning the tasks. Other networks, such as JEDI-net \cite{Moreno:2019bmu}, Point Cloud Transformer \cite{Mikuni:2021pou}, Lorentz Net \cite{Gong:2022lye}, and PELICAN \cite{Bogatskiy:2022czk}, are also utilized for particle cloud analysis, and we plan to compare our results with these models.

Particle Transformer \cite{Qu:2022mxj} is a Transformer-based particle cloud model. 
Transformers were initially proposed as sequence-to-sequence models for machine translation \cite{vaswani2017attention} and modified for collider purposes in \cite{Qu:2022mxj}. 
At the core of the particle transformer, the attention mechanism which enables  the model to focus selectively on different parts of input data, assigning varying degrees of importance to each component. 
Since the attention mechanism computes the attention weights of each particle to all other particles in the dataset, these weights remain unchanged even if the order of the input tokens is permuted. The transformer models have shown the best performance for collider analysis. However, they suffer from high model complexity and require extensive run time.

This paper presents a novel permutation-invariant network that capitalizes on superior performance with reduced runtime. This can be achieved by analyzing subjet and jet constituent information via cross-attention heads and a "Mixer network" consisting of multi-layer perceptrons(MLPs).
 
The utilization of subjet information in jet clustering has already appeared in the original ideas of jet classifications\cite{Butterworth:2008iy}, aiming to identify the cluster location inside the jet for jet classification. 

The subjet momentum is an IRC-safe quantity that matches the hard parton momentum originating jet.  TMixer network integrates two MLPs to capture local and global information about the event. Like transformers, these MLPs amalgamate particle and feature tokens from the dataset, facilitating efficient structure learning.
The first MLP combines all particle tokens in the dataset, with weight sharing enabling the model to learn the dependencies among each particle and all others. Meanwhile, the second MLP mixes the features of all particle tokens, resulting in a dataset with the same dimensionality as the inputs. 
The concept of the mixer layer, initially introduced for image classification in \cite{tolstikhin2021mlp}, is central to our model. However,  the MLP part of the mixer layer is not inherently permutation invariant and relies on the order of input particles within the set. 

We utilize a cross-attention mechanism that 
analyzes pre-clustered subjet data to achieve both permutation invariance and model performance. 
Adjusting cross-attention heads to analyze jet kinematics and jet constituents has already been introduced in \cite{Hammad:2023sbd,Buhmann:2023acn}, where global event kinematics and constituents of the jets are first processed individually by self-attention heads, and subsequently combined by cross-attention. In this setup,  jet constituents become the keys, and global event kinematics become queries and values.  The scaled cross-attention score becomes the matrix of the dot product between the query and key matrices, which is trained to highlight the quantity significant for the event classification.  In short, jet constituent information valuable for event classification is used to transform jet momenta.  

\begin{figure*}[th!]
\includegraphics[scale=0.25]{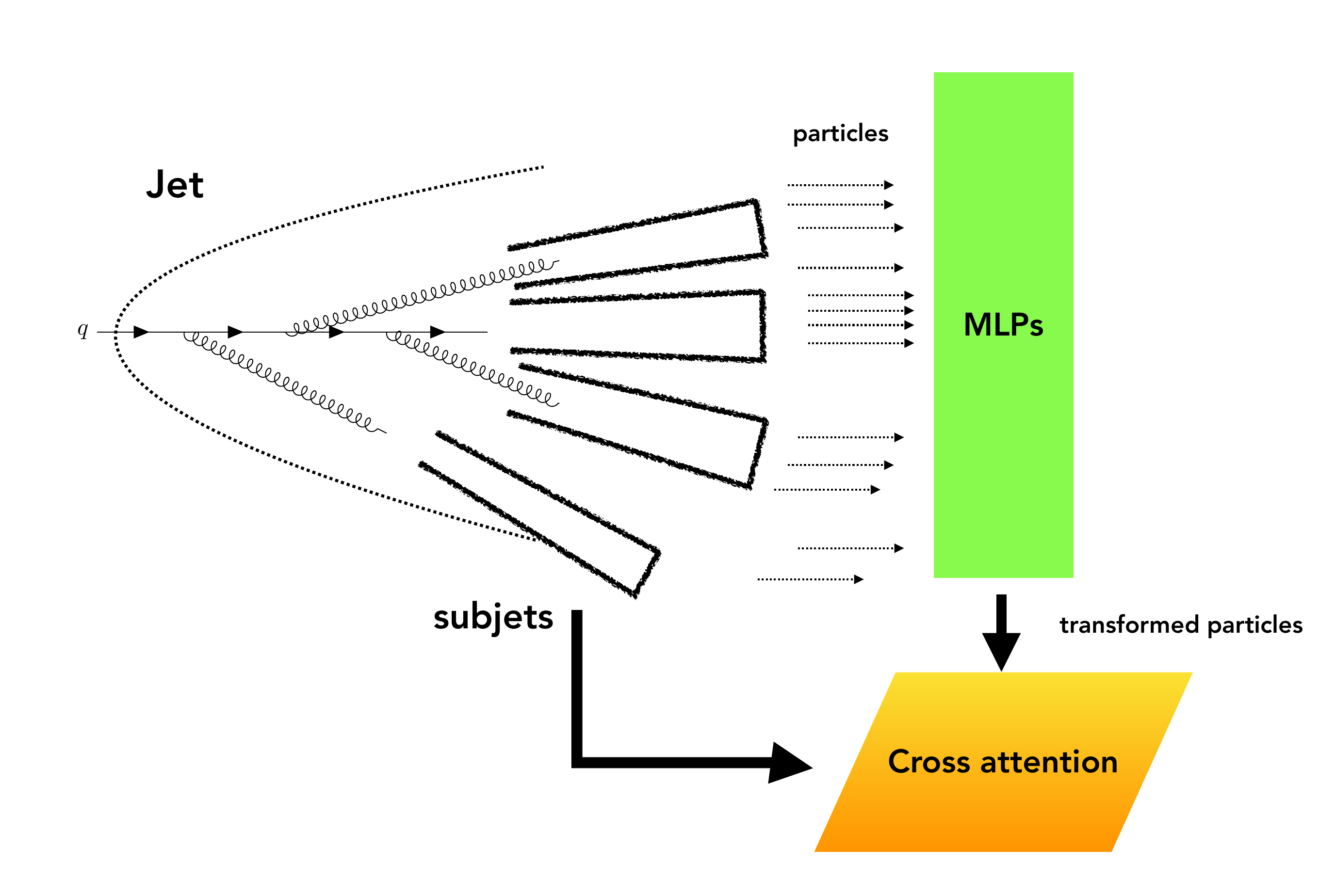}
    \caption{Schematical figure of the mixer layer in the Mixer network. A hard parton from the hard process goes through the parton shower, creating subjets. The information of jet constituents is processed by two MLPs, then analyzed together with subjet information via the cross attention layer.}
    \label{fig:mixernet}
\end{figure*}

In this paper, similar to the previous setup, we apply the cross-attention for jet substructure analysis; the key matrix is constructed from subjet momenta,
while jet constituents become the query and the value matrices.  These cluster momenta are closely related to the parton momenta from the hard process or resolved emissions from the parton shower. The applied cross-attention heads weigh the relation between clusters and the jet constituents contributing to the event classification. Moreover, the network structure is well-suited for estimating the probability of hadron collider processes. This is because the final state particle distribution of the hadron collider may be described schematically in the factorization limit as follows, 
\begin{equation}
P(x)= \Pi_i P_s(\{x\}\vert y_i)  P_h(\{y_i\})\label{eq:fac}
\end{equation}
where the $P_s$ is the probability of the hardons $\{x\}$ conditioned by a parent parton information $\{y\}$ while $P_h$ is the parton distribution of the hard process involving $\{y\}$. The hadron distribution in the jet correlates with hard parton momenta through the cluster momenta. Therefore, a model based on the jet-to-jet constituent cross-attention is promising in representing the QCD process. See the schematical figure of relations between the parton shower process and hadronization in Fig.\ref{fig:mixernet}. 

This paper is organized as follows. In section \ref{sec:2}, we describe the network structure constituted with mixer layers for the constituents and cross-attention network to incorporate subjet information. We pay particular attention to assure the resultant network permutation invariant. We describe our top and QCD sample for network  validation in section \ref{sec:3}. 

In section \ref{sec:4}, we describe the different sub-jets clustering methods studied in this paper.   
We utilize standard sequential recombination algorithms such as  Cambridge-Aachen (CA) \cite{Dokshitzer:1997in} and Anti-kt \cite{Cacciari:2008gp}, which typically require a fixed radius parameter for clustered subjets. 

Because the fixed size radius parameter could be a limitation in capturing the jet structure and require hyperparameter tuning,  we employ 
Hierarchical Density-Based Spatial Clustering of Applications with Noise (HDBSCAN) \cite{10.1007/978-3-642-37456-2_14} which does not depend on the radius parameter and use a proper distance metric to ensure Infra-Red and Collinear (IRC) safe clustering. 

The result is given in section  \ref{sec:5}. In section \ref{sec:6}, 
we explain the network output using different interpretable methods, including centred kernel alignment and attention maps to highlight network efficacy in collider analysis tasks.

\section{Network architecture}
\label{sec:2}

In this section, we explain the structure of our network. 
As we already stressed in the introduction, the core of our networks is a simple mixer layer integrated with the subjet information by cross-attention so that the network maintains the hierarchy between low- and high-scale physics.

The proposed network comprises distinct layers: an input layer, a mixer layer, an aggregation layer and a final fully connected (FC) layer. The permutation invariance of the network is ensured by the aggregation layer and cross-attention heads within the mixer layer. 

\begin{figure*}[th!]
    \centering
    \includegraphics[width=\textwidth]{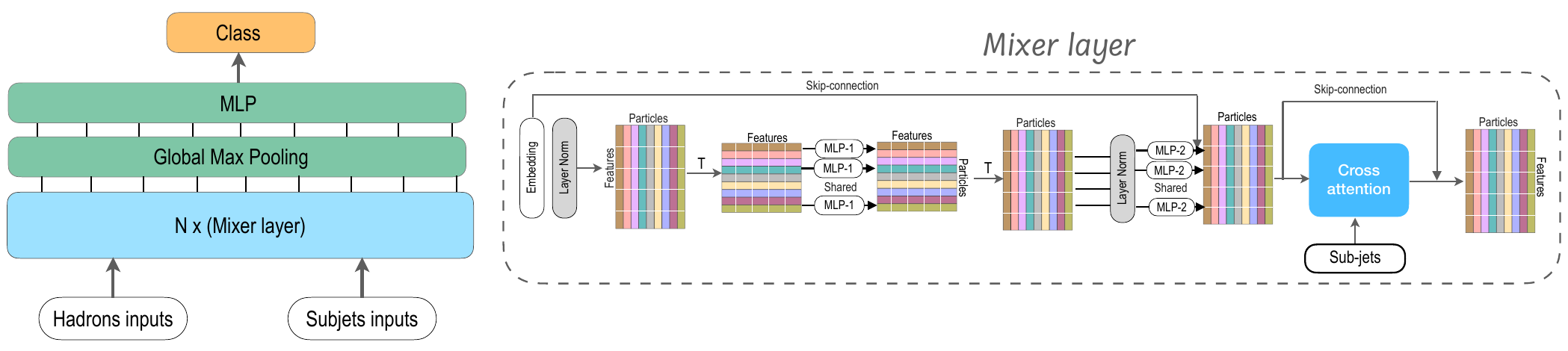}
    \caption{Left: full network structure. Right: structure of the mixer layer block. Each MLP within the mixer layer shares its weights and comprises two FC layers and a Gaussian Error Linear Units (GELU) layer.}
    \label{fig:network}
\end{figure*}

The core of the network is the mixer layer, which consists of two components, two MLPs  and cross-attention heads and discussed in subsection II. A. 
The first MLP acts on each particle in the cloud individually, while the second one acts on each feature of the mixed particles after transposing the dataset. 
The MLP shares weights across the processing layers, ensuring that all particle and feature tokens obey the same transformation 
(See figure \ref{fig:network}).  This allows the network to learn a unified representation among different features. 

Input data to the mixer layer
passes sequentially through two MLPs consisting of densely connected neural network layers. It is then passed to the cross-attention heads along with the subjet dataset.
Note that the two MLPs operate similarly to transformer models with self-attention heads, combining particles and their corresponding features across the entire dataset. This enables the extraction of local and global structural information within the event. A side effect is that the MLPs have smaller tunable parameters to express the complex structure of the event compared to the other particle cloud models such as particle Net or transformers.

To compensate for the less expressivity due to smaller parameters of the Mixer network, 
we introduce a second input dataset containing 
subjets to discern the substructure of top and QCD jets. The details of the subjet clustering methods are not essential for the network description and are discussed in Section 4.  The additional dataset, together with the output of the MLPs,  are analyzed by the network using multi-head cross-attention and described in subsection II. B. 
The mixer layer preserves the dimension of the input dataset and can be repeated for better performance with more complex data. The network structure is shown in Fig. \ref{fig:network}. 

To further ensure the permutation invariance of the network, mixer layers are followed by a global Max-Pooling layer. An additional FC layer is added before the output layer with two neurons.
\subsection{MLP mixer}
At the heart of the MLP Mixer lies its features mixing mechanism. 
It begins with feature mixing by transposing the particle and feature axes, and then it continues with particle MLP mixing so that the input data is mixed in the feature and particle axes.  Consider the input data set $X_{(i,j)}$ in which $i$ and $j$ run over the particle tokens and their features, respectively. The input data is first passed by a linear FC layer to map the features to higher dimensions. The first MLP acts on the features of each particle token individually mixing them up into new features as:
\begin{equation}
    Y_{i,j} = X_{i,j} + [\left( W_2 \sigma W_1 (\text{LayerNorm}({\bf X}
    )^T)   \right)^T]_{ i,j} \,,
\end{equation}
where $W_1, W_2$ are the weight matrix of the first MLP FC layers, and $\sigma$ is the activation function that acts on each component. The output is then passed to the second MLP to mix the particle tokens as follows
\begin{equation}
    X^\prime_{i,j} = Y_{i,j} + \left( W_4 \sigma W_3 (\text{LayerNorm}(Y_{i,j})))\right)\,,
\end{equation}
with $W_3,W_4$ are the weights of the second MLP FC layers and $\sigma$ is the activation function. The skip connection ensures that the input ($X_{i,j}$)  and output ($X^\prime_{i,j}$) are strongly correlated. 
The output $X^\prime_{i,j}$, together with the clustered subjets dataset, are the inputs to the multi-heads cross-attention.  
\subsection{Cross-attention}
The advantage of cross-attention lies in its ability to capture relationships and dependencies between different input sequences or modalities. Unlike self-attention, which operates within a single sequence, cross-attention allows a neural network to incorporate information from one sequence into the processing of another by giving a structure that is suitable for
understanding interactions between different classes of the input data. 

In the context of our paper, the subjet momentum and jet constituents have different modalities. 
In principle, the structure of jet clustering can be learned from the jet constituent information. 
However, having the subjet clustering information as a separate dataset, the network can easily assign jet constituents to the subjet through their location information, allowing the network to concentrate on the structure of each subjet.

Considering the input to the multi-head cross-attention as $X^\prime_{i,j}$, which represents the output of the second MLP, along with the subjet information $S_{n,m}$.  A linear FC layer passes $X^\prime$ to generate the weight matrix for the Query matrix, while  other FC layers  generate the weights matrices for the Key and Value from $S$ as 
\begin{equation}
\begin{split}
    &Q^{i\times j} = X^{\prime i\times j}\cdot W_Q^{j\times j}\,, \hspace{4mm} K^{n\times j} = S^{n\times m}\cdot W_K^{m\times j} \,, \\
    &\hspace{20mm} V^{n\times j} = S^{n\times m}\cdot W_V^{m\times j} \,,
    \end{split}
\end{equation}
where $Q, K$ and $V$ are the query, key and value matrices, respectively, and will be used to compute the attention of the dataset. The superscripts indicate the dimension. The scaled dot product attention score is defined as
\begin{equation}
\begin{split}
    \alpha^{i\times n} &= \text{softmax}\left(\frac{Q^{i\times j} \cdot (K^{n\times j})^T}{\sqrt{d}} \right) \\
    &=\frac{\text{exp}(Q^{i\times j}\cdot K^{j\times n}/\sqrt{d})}{\sum_i\text{exp}(Q^{i\times j}\cdot K^{j\times n}/\sqrt{d})}\,.
\end{split}
\end{equation}
The resulting attention weight matrix has the dimensions of (particle tokens $\times$ subjet tokens) and assigns each particle token to its subjets. The attention output is obtained as 
\begin{equation}
    \mathcal{Z}^{i\times j} = \alpha^{i\times n}\cdot V^{n\times j}\,.
\end{equation}
The attention output matrix has the same dimension as the first input dataset $X$. It illuminates the importance of each particle token once assigned to the corresponding subjets and the entire dataset. This assignment of each particle token to its corresponding subjets enables the attention output matrix to emphasize the crucial particle tokens for learning the substructure of the top jet. In this case, the attention output matrix exhibits a different structure between top and QCD jets, as our discussion in section \ref{sec:6}.

The process described above is repeated for each attention head, resulting in multiple attention outputs for each token.
Finally, the attention outputs from all heads are combined and projected into a single representation. This combined representation captures different aspects of the input data learned by each attention head. The output of the multi-head cross-attention has the form
\begin{equation}
    \mathcal{O}^{i\times j} = \text{concat}\left(\mathcal{Z}^{i\times j}_1,\mathcal{Z}^{i\times j}_2 \cdots \mathcal{Z}^{i\times j}_n \right) W^{(n*j\times j)}\,,
\end{equation}
with $W^{(n*j\times j)}$ is the learnable linear transformation matrix to retain the dimensions of the input dataset. 
Attention output is used to scale the input data set via a skip connection as
\begin{equation}
    \widetilde{X}^{i\times j} = X^{i\times j} + \mathcal{O}^{i\times j}\,.
\end{equation}
The transformed dataset $\widetilde{X}$ signifies the importance of each element relative to all elements within the set. While the attention output integrates input and feature tokens, the skip connection preserves the correlation to the original input dataset. Moreover, it preserves the dimensions of  $X_{i,j}$.

Ultimately, the transformed dataset undergoes processing by a global Max-Pooling layer, identifying the particle token with the highest score. 
The global max pooling operates as the following
\begin{equation}
    Y_j = \text{Max}_{i=0}^{k-1}  \widetilde{X}^{i\times j}\,,
\end{equation}
where $k$ is the number of the particle tokens in the dataset. 
While any symmetric aggregation function could be utilized to maintain the network's permutation invariance, but we found that Max-Pooling has the best performance \cite{qi2017pointnet}.

The output is then passed to a FC layer with ReLU activation and an output layer with two neurons. The final output score  
has the form 
\begin{equation}
    \hat{Y} = \text{Softmax} \left[ W_6\ (\text{ReLU }(W_5 \ Y_j)) \right]\,,
\end{equation}
which encodes the probability of the input event to be 
signal or background event.
\subsection{The role of cross-attention for collider physics}

The cross-attention network is suited to study the correlation between hard partons and hadrons in the events. 
Considering a hard process leading $N$ final jet, the factorization picture connects the parton distribution to the hadron distribution as follows \cite{Walsh:2011fz}, 
\begin{equation}
\sigma(p p \rightarrow a, b \rightarrow N \text{jets}) \sim  H_N \left[B_a B_b  \prod^N_{k=1} J_k\right]\otimes S_N \,,\label{eq:SC}
\end{equation}
where $H_N$ express the hard scattering cross section, $B_{a}$ and $B_{b}$ is the beam function; $J$
express the collinear evolution of hard partons from the hard scattering,   and the soft function $S_N$ expresses the soft radiations. The formula suggests that the soft hadron distributions in a jet are conditioned by the hard process $H_N$, the parton evolutions, and the hadronization processes that connect all partons.   

Due to the correlation between parton momenta and jet momenta, the QCD process may schematically be expressed as 
 
\begin{equation}
\prod P_s( \{x_k\}\vert \{J_i\} )P_h( \{J_i\})  \,,
\label{eq:facii}
\end{equation}
where $P_s$ is the hadron distributions in the jet, conditioned by the jet features, and $P_h$ is the distribution of jets, which approximately express $H_N \prod_k J_k$. 
Note that $P_s$ is conditioned by all jets in the events due to the effect of $S_N$ in Eq.\ref{eq:fac}. Eq.\ref{eq:fac} is a much simpler approximation, which assumes hadrons arising from a single parton. 

In our network, the cross-attention score is computed as  $\alpha=QK^T$, which is the product of the output from the mixer layer and the subjet information. Therefore, the network is strongly directed to study the structure given by  Eq\ref{eq:facii}. Taking the correlation between all subjets and all constituents to take care of $S_N$ factor in our network. 
Note that the splitting between $P_h$ and  $P_s$ has ambiguity on the choice of jet radius parameter $R$. If one takes smaller $R$, the number of subjets increases by splitting subjets. In Eq \ref{eq:SC}, this corresponds to the change of the resolving scale of the parton shower. The radius $R$ is an ad-hoc parameter of our network.  The proper choice of the radius parameter $R$ for our event sample and method, which does not rely on the radius parameter $R$, will be discussed in section IV. 

\section{Top tagging dataset}
\label{sec:3}
Top tagging, namely the identification of jets originating from hadronically decaying top quarks, is crucial in searches for new physics at the LHC. To assess the effectiveness of the proposed network, we utilize the top tagging dataset \cite{Butter:2017cot}. Jets in this dataset are generated in the centre of mass energy $\sqrt{s}=14$ TeV using Pythia8 \cite{Bierlich:2022pfr}. Delphes \cite{deFavereau:2013fsa} is used for fast detector simulation. The simulation does not account for multiple parton interactions or pileup effects. The jets are clustered from Delphes E-Flow objects using the Anti-kt algorithm with a cone of radius $R=0.8$.
Jets 
with transverse momentum $p_T\in [550,650]$ GeV and pseudo rapidity $|\eta| < 2.$ are considered. 
For top events, the event should contain the jets that match the top quark, namely, a jet within $\Delta R = 0.8$ from a hadronically decaying top quark and also all the three quarks from the top decay are within  $\Delta R=0.8$ from the jet axis. 

The QCD dijet process is considered as the background. 

The data set contains 1 million $t\bar{t}$ events and 1 million QCD dijet events. 
We adhere to the official split for training 1.2M event, validation 400k event, and testing 400k event. The data sample has been widely used in the previous literature, making it easy to compare the network performance with the others. One drawback of using this sample is the effective sample imbalance around the top mass region; the top sample peaks around 170 GeV while the QCD sample peaks near zero; in other words, the overlap between the top sample and the QCD sample is poor, making it difficult to compare the fine difference among the high-performance networks. 

Up to $200$ constituent particles (hadrons) are retained for each jet in the dataset, with the 4-momenta $(px, py, pz, E)$ of each particle. 
From this dataset, we construct the first input dataset with up to $100$ $p_T$ ordered jet constituents with seven features for each particle as:
\begin{itemize}
    \item $\Delta\eta = \eta - \eta_{jet}$, where $\eta$ ($\eta_{jet}$) is the pseudorapidity of each constituent (jet).
    \item $\Delta\phi =  \phi - \phi_{jet}$,  where $\phi$ ($\phi_{jet}$) is the azimuth angle of each constituent (jet).
    \item $\Delta R = \sqrt{\Delta\eta^2+\Delta\phi^2}$, which represents the angular distance of each constituent from the jet axis.
    \item $\log(p_T)$, transverse momentum of each constituent in GeV.
    \item $\log(E)$, energy of each constituent in GeV.
    \item $\log(p_T/p_{T_{jet}})$, normalized transverse momentum of each constituent in GeV.
    \item $\log(E/E_{jet})$, normalized energy of each constituent in GeV.
\end{itemize}
The first input dataset has the dimension $(100,7)$ with the first and second numbers referring to the maximum number of jet constituents and the features, respectively. Any events with fewer jet constituents are padded with zeros to maintain the uniformity of the data size. 
The second input dataset is the subjet information of the jets with the size of $(15,7)$
 
The first index denotes the maximum number of subjets. Again, if the number of subjets is less than 15, the remaining arrays are padded with zeros. 
The second index denotes the subjet features that are the same as the feature of jet constituents.

\section{subjets clustering}
\label{sec:4}
In jet classification tasks using ML, 
the local structure of the dataset may be extracted from its nearest neighbours.  Network models can extract the data patterns to tag jets by finding the useful correlation among the jet constituents. \cite{Li:2022xfc}. 
However, if we ask the network to analyze the correlation among the jet constituents from scratch, the required computational resources increase significantly. Instead, one can incorporate the information by adding some feature representation. For this purpose, we introduce the network assisted by the subject information, which encodes the local structure of the jet.

Unlike the non-parametric approach using ML models, the subjet depends on the choice of the jet clustering algorithms. 
In the realm of hadron collider physics, the preferred choice of jet clustering is sequential recombination algorithms:  Cambridge-Aachen (CA), kT, or anti-kt. 
These algorithms have several advantageous properties with minimal parameter adjustments provided by a computationally efficient package,  FastJet \cite{Cacciari:2011ma}.
; they are IRC safe and have the flexibility to capture various jet natures in different environments effectively; for example, CA is suited to compare the data with QCD calculation, while anti-kt is used to reconstruct the jets with underlying events. 

These algorithms operate on a recursive or iterative basis and are agglomeration. 
Agglomerative clustering is a bottom-up hierarchical method where each data point starts its cluster.
The clustering process involves computing pairwise distances between clusters called pseudojet, merging the closest ones, and updating the distance matrix until it exceeds a threshold distance. Ensuring the infrared safety of created jets entails an early combination of pairs of particles resulting from soft and collinear emissions during the clustering process. It should be noted that the clustering sequence carries the important hint of parton shower process, and the graph neutral network utilizing the clustering sequence, LundNet\cite{Dreyer:2020brq}, achieves high performance. 

Although the success of jet clustering algorithms, 
it has ad-hoc parameter $R$ that determines the cluster size, 
indirectly affecting the number of constructed jets \cite{Cerro:2021abp,Mukhopadhyaya:2023rsb}. 

HDBSCAN is a clustering based on hierarchical density estimate, and the algorithm is not associated with any distance parameter $R$. HDBSCAN defines clusters adaptively by leveraging the density of data points within proximity, and does not have a predetermined jet radius parameter. 
It identifies meaningful clusters, and HDBSCAN distinguishes outliers and noise so that soft particles in low density regions are considered noise points and left unclustered, thus providing a comprehensive perspective on the data structure. 

In the rest of this paper, we consider the three clustering algorithms to prepare the subjet dataset. Also, we test the impact of each algorithm on the classification performance of the network.

\subsection{Clustering with radius parameter}

The CA and anti-kt algorithm is commonly used in jet clustering. It takes the nearest neighbour method, namely 

for the minimum distance pair $i$ and $j$ with distance $d_{ij}$, one replaces $i$ and $j$ with a new object called "pseudojet" with momentum $p_i+ p_j$. If the smallest distance is a distance to the beam $d_{iB}$,  the particle $i$ is removed from the list. This procedure is repeated  until the smallest $d_{ij}$ or $d_{iB}$ is above some threshold $d_{cut}$

\begin{figure}[th!]
    \centering
    \includegraphics[scale=0.27]{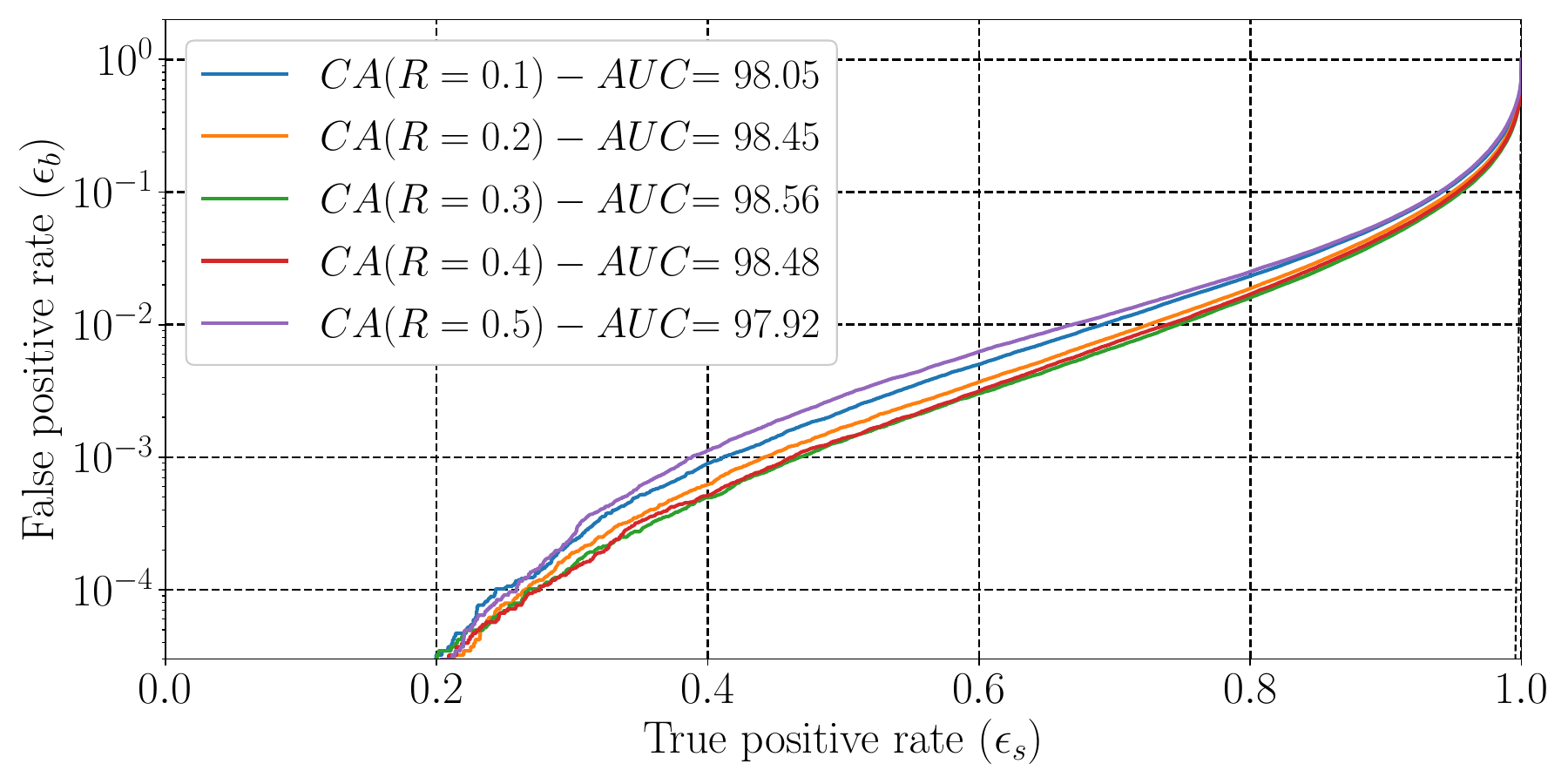}
    \caption{Receiver Operating Characteristic (ROC) curves for the mixer network trained on a secondary subjets dataset clustered with CA with different $R$ values.}
    \label{fig:2}
\end{figure}

The CA algorithm comes with a pair distance measure as 
\begin{equation}
d_{i,j}  = \frac{\Delta R^2_{i,j}}{R^2}\,, \hspace{4mm} d_{iB} =1 \,,
\end{equation}
where $\Delta R_{ij} = \sqrt{(\eta_i-\eta_j)^2+(\phi_i-\phi_j)^2}$ and $d_{iB}$ is the distance of a parton $i$ from the beam. 

On the other hand, the anti-kt algorithm  \cite{Cacciari:2008gp} comes with a pair distance measure as 
\begin{equation}
    d_{i,j}  =min(\frac{1}{p^2_i},\frac{1}{p^2_j}) \frac{\Delta R^2_{i,j}}{R^2}\,, \hspace{4mm} d_{iB} =\frac{1}{p^2_i} \,.
\end{equation}
Hard anti-kt jets have circular shapes on the $\eta-\phi$ plane and look like jets in a cone algorithm \cite{Cacciari:2008gp}

To examine the impact of radius parameter $R$ on the model classification performance, we tested the network performance for different values of $R$,  ranging from $0.1$ to $0.5$ using the CA algorithm.
Figure \ref{fig:2} illustrates the performance of the mixer network for classifying top and QCD jets, evaluated using the Area Under the ROC (AUC) metric.  
Optimal classification performance is observed at $R=0.3$, with deviations from this value resulting in decreased accuracy. This behaviour is expected, as increasing/decreasing $R$ may dilute the subjets structure for the top events.
\subsection{Dynamic radius clustering (HDBSCAN)}
HDBSCAN represents a sophisticated extension of traditional clustering algorithms of point data.  
It introduces a hierarchical approach that identifies clusters of varying densities. 
Unlike conventional clustering algorithms that necessitate prior specification of cluster count, HDBSCAN autonomously discerns clusters of varying sizes and shapes. This adaptability renders HDBSCAN particularly well suited to capture meaningful geometrical information about the substructure of the top jet. The algorithm is based on the k-nearest neighbourhood method and is as fast as Fasjet package. 

HDBSCAN begins by computing the mutual reachability distance between particles in the event, creating a reachability distance matrix.  
Then, HDBSCAN constructs a minimum spanning tree from this matrix and identifies the particles corresponding to the peaks in the density of the minimum spanning tree. These particles become the initial cluster centres. Next, the algorithm performs a hierarchical clustering of the particles using a variation of single linkage clustering. Finally, it assigns each particle to a cluster based on the hierarchical clustering structure, leaving particles in low-density areas as noise points. 
 
\begin{figure}[th!]
    \centering
    \includegraphics[scale=0.23]{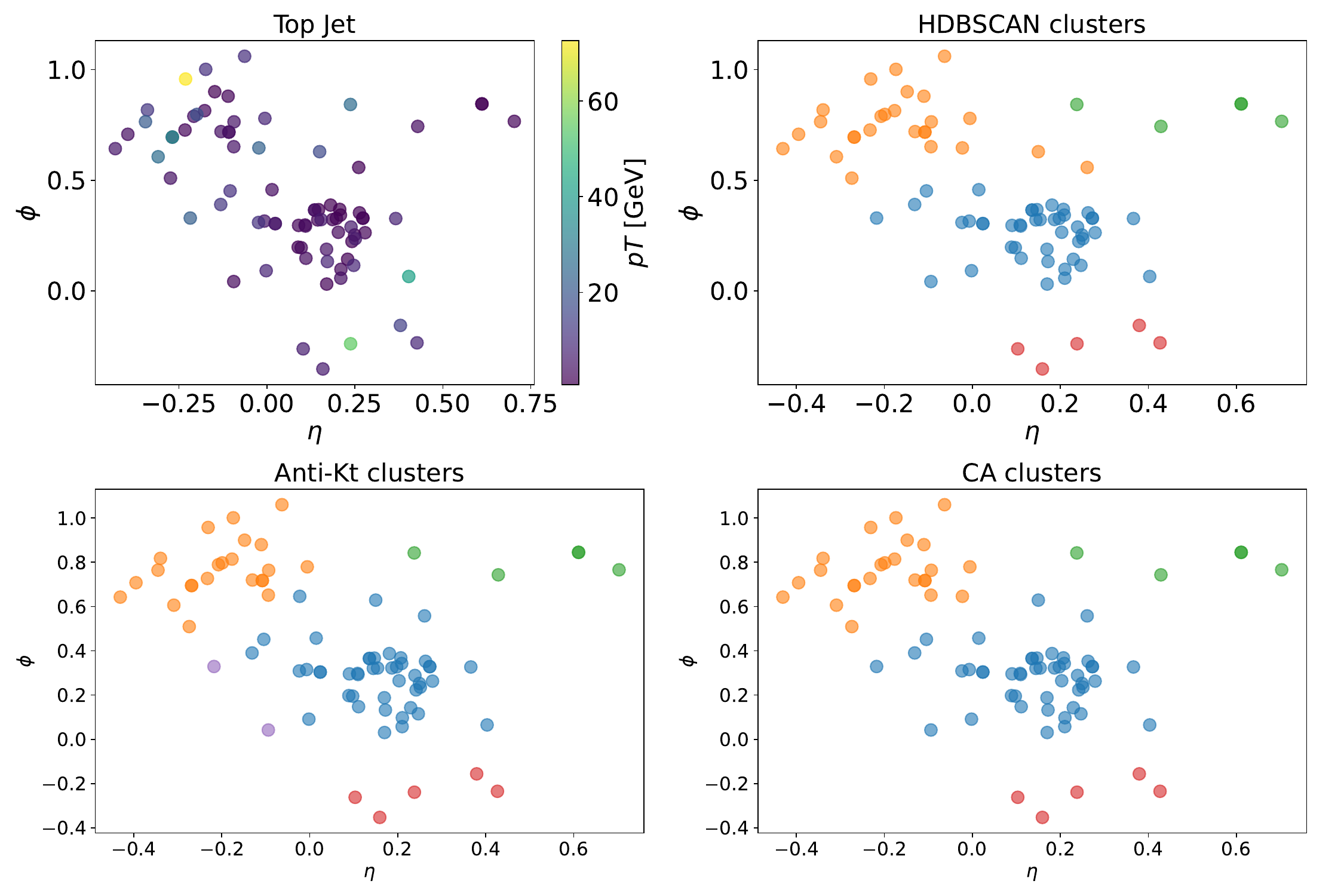}
    \caption{Example event for top jet clustering. Left top: distribution of all top jet constituents in the $\eta-\phi$ plane. Right top: The results of the clustering of the subjets using HDBSCAN. Left bottom: The results of the clustering of the subjets using Anti-Kt with $R=0.3$. Right bottom: The results of the clustering of the subjets using CA with $R=0.3$. The points with the same colour represent the same cluster members.}
    \label{fig:3}
\end{figure}
In Fig. \ref{fig:3}, we show an example of subjets clustering using  the different clustering algorithms, HDBSCAN, Anti-Kt and CA for top jet example. The top left plot shows the distribution of all top jet constituents in the $\eta-\phi$ plane while the colour bar indicates the transverse momentum. The top right plot shows the resulting three subjet clusters in red, green, and orange colours. The four green coloured particles are left unclustered because they are relatively in low density areas. The bottom left (right)  plot shows the resulting five (four) subjet clusters when using Anti-Kt (CA) with radius $R=0.3$. 
Due to the unclustered particles, there is a mismatch of the original jet momentum and the sum of the reclustered subjet momenta and the jet momentum; for HDBSCAN clustering on top and QCD samples,  an average value of $(\sum P_{\rm sub})/P_{\rm jet}$$ = 0.983$ and $75.97\%$ of the events has $\sum P_{\rm sub}\ge 95\% P_{\rm jet}$, with $P_{\rm sub}, P_{\rm jet}$ are the subjet and jet momentum\footnote{A common problem with most of the ML algorithms is the Pileup effect, but modern machine
learning algorithms like PUMML \cite{Komiske:2017ubm}  and others  can effectively eliminate the pileup
effect.}.

Subjets are clustered as follows: 

It starts by calculating the distance from the $k^{\text{th}}$  nearest neighbour (the core distance)  for all the particles. 
This  $k^{\text{th}}$ is the discrete hyper-parameter of HDBSCAN. A particle $p$ is a core particle if the distance to all its nearest neighbours is less than the distance to other core particles. We modify the reachability distance with particle transverse momentum as follows; 
 
\begin{equation}
    d_{i,j}  = \sqrt{(\eta_i - \eta_j)^2+(\phi_i - \phi_j)^2}\times S_{i,j}\,,
    \label{eq:1}
\end{equation}
where $\eta,\phi$ are the pseudorapidity and azimuth angle of the considered particle. $S_{i,j}$  is a singularity factor to ensure that softer particles are  combined earlier to the nearest cluster as discussed in \cite{Cerro:2021abp}
\begin{equation}
    S_{i,j} = 1- \frac{k}{k-\text{min}(Pt_i,Pt_j)\sqrt{(\eta_i - \eta_j)^2+(\phi_i - \phi_j)^2}}\,,
\end{equation}
where $k$ is a small number taken to be $0.0001$. The singularity factor is  $0$ for soft and collimated particles and 1 for well-isolated particles. Therefore, this distance measure interpolates the distance measure of the $k_T$ and CA distance measures. Note that this definition could make the algorithm sensitive to the underlying events, while it helps to network actively capture the geometry of the soft particle distributions. 

Once the core distances are defined, the density of the particles in the $\eta-\phi$ plane is defined via a mutual reachability distance measure ${\text{r}}$, which is defined as 
\begin{eqnarray}
    d_{\text{r}}&&(x_i,x_j) = \cr 
    &&\scalemath{0.9}{\begin{cases} 
    \text{max}\left(\text{core}_k(x_i),\text{core}_k(x_j),d(x_i,x_j)\right) \hspace{4mm} x_i\neq x_j\, ,\\ 
    0 \hspace{10mm} x_i =  x_j \,, \end{cases}}
\end{eqnarray}
where $\text{cor}_k(x_i), \text{cor}_k(x_j)$ are the core distance of $x_i$ and $x_j$ respectively from their $k^\text{th}$ nearest neighbour, and $d(x_i,x_j)$ is the distance between the two particles, as defined in Eq. \ref{eq:1}. 
Note that the distance between the particles is replaced with the core distance if the particle is in a sparse area; therefore, the particles in the sparse area tend to be pushed away from the other clusters. 

\begin{figure}[th!]
    \centering
    \includegraphics[scale=0.23]{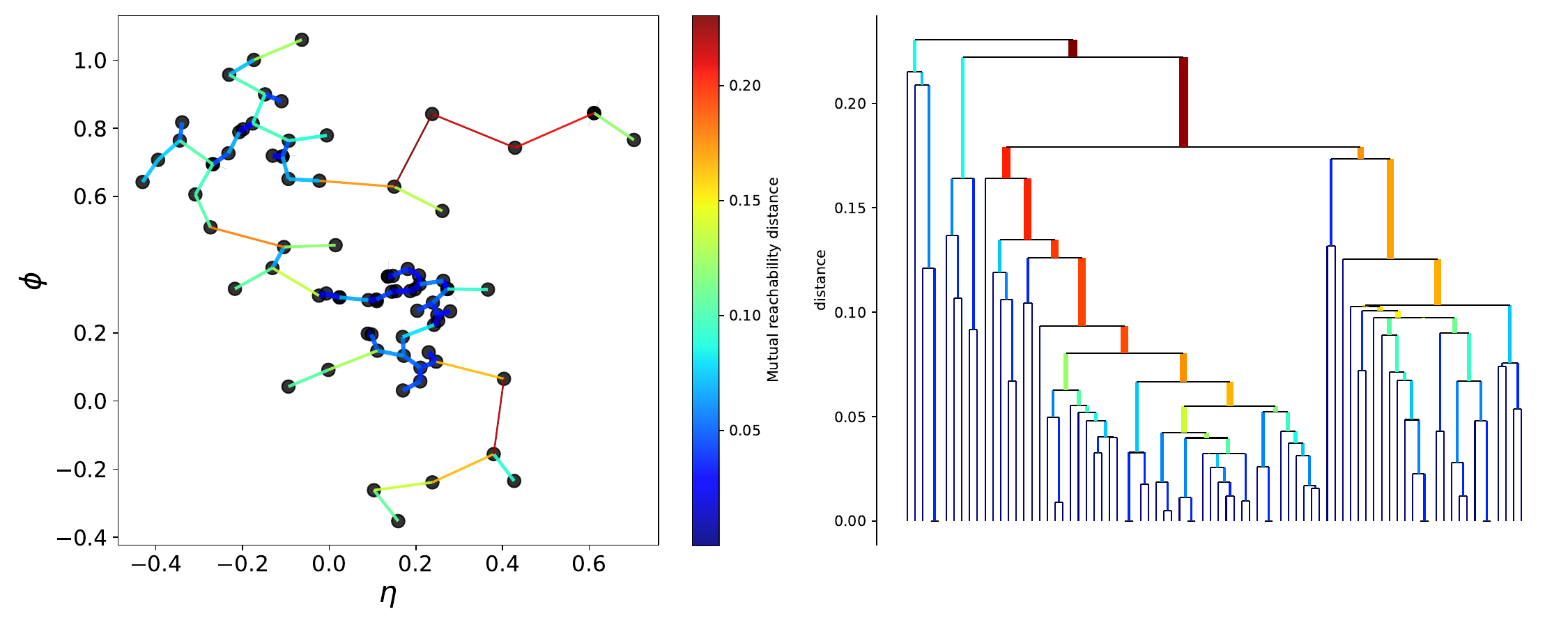}
    \caption{Right: the minimum spanning tree for top example event. The colour bar represents the mutual reachability distance between the reachable particles. Left: dendrogram for Single Linkage Tree.}
    \label{fig:4}
\end{figure}

For hierarchical clustering, one constructs a graph with nodes representing particles and edges connecting nodes with the weight of the mutual reachability distance.  The graph called the minimum spanning tree is built by considering one edge at a time, always adding the lowest weight edge that connects the current tree to a vertex not yet in the tree. In the left plot of Fig. \ref{fig:4}, we show the constructed minimum spanning tree of an example top jet event in Fig. \ref{fig:3}. This figure clarifies that the unclustered particles, the four blue points in Fig. \ref{fig:3},  assigned larger mutual reachability distance shown in red lines. 

Having the minimal spanning tree constructed, the next step is to convert that into the hierarchy of connected components. This can be done by sorting the edges of the tree by distance 
for which particles with smaller distances are merged first and then iterate through, creating a new merged cluster for each edge and ending up with a single linkage tree. In Figure \ref{fig:4}, right plot, we show a dendrogram for the reconstructed hierarchical structure from the minimum spanning tree. The length of the branches in the dendrogram represents the distance between clusters and/or clustered particles in the event. 

Once we obtain the dendrogram, we go back to the clustering sequence to the small distance pairs to identify stable clusters. As depicted in the single linkage tree above, it is common to observe a cluster of one or two points separate from the main cluster. 
Instead of interpreting this event as the split of a cluster, 
we perceive it as a single enduring cluster experiencing a reduction of points. For this purpose, we require the minimum number of points $m$ for a single cluster. 
It is another hyperparameter of the algorithm and we fix it to $m= 5$. 
If each child cluster contains particles less than $m$ at a cluster split, this split is considered spurious, and the particles will be removed from the parent cluster. Conversely, if the split is into two clusters, each at least as large as the minimum cluster size, we consider a true cluster split and let that split persist in the tree. In this manner, the tree size shrinks to smaller clusters with larger stability. 

To measure the stability of the condensed tree and choose clusters that persist for long during the splitting, a density parameter is introduced as $\lambda=\frac{1}{d}$, where $d$ is the distance represented in the vertical axis in Fig. \ref{fig:4} (right). The stability of each cluster can be defined as the sum of the $\lambda$  for particles in the cluster; 
\begin{equation}
    \text{Stability}(j) = \sum_{p\in j} \left(\lambda_p - \lambda_j \right)\,.
\end{equation}

The final cluster of HDBSCAN is the trees with significant stability. 
To select the final clusters, we start by declaring all nodes as selected clusters. Then, traverse the tree toward larger $\lambda$.  
If the combined stabilities of child clusters exceed that of the parent cluster, we update the parent's stability to equal the sum of its children's stabilities. Conversely, if the parent's stability surpasses the sum of its children, designate the parent cluster as a selected cluster and deselect all its descendants. Upon reaching the root node, consider the selected clusters as our final subjet. 

In the selection process, there is no distance nor the (sum of)  the momenta involved. In the case of CA or Anti kT algorithms, the momentum is always updated to the sum of the momenta of the parents, where the most energetic constituent dominates the direction.  One may imagine that the clustering pattern can completely differ from the CA or anti-kT. However, it is natural that high energy particles create collinear particles, Therefore,  the direction of the high energy cluster and the high density location are strongly correlated. 

To use the HDBSCAN for jet analysis, it is necessary to calibrate the relation between parton momenta and momenta of the clusters. For example, the relation of the CA subjet and HDBSCAN cluster should depend on the colour of the parent parton and its colour connection. 
The effect of the colour connection might be interesting to study because HDBSCAN is more sensitive to the distribution of soft particle constituents. 

\begin{figure}[th!]
    \centering
    \includegraphics[scale=0.17]{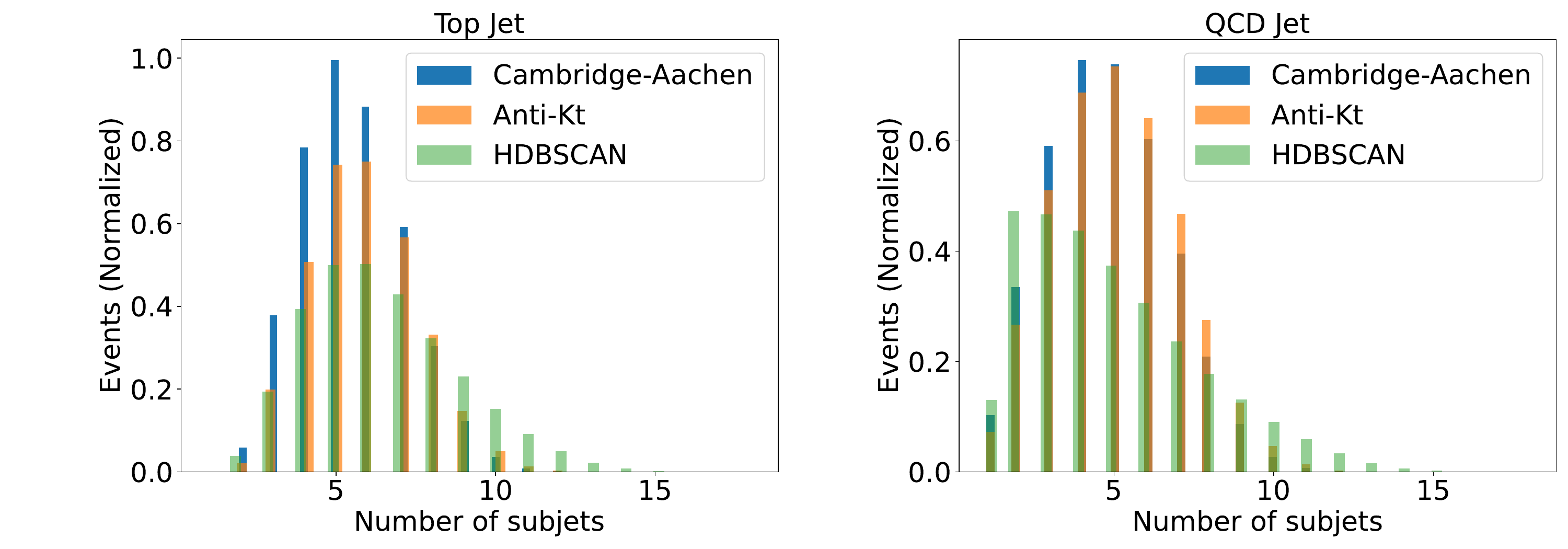}
    \caption{Number of subjets in the Top jet events (left) and QCD jet events (right) clustered with Cambridge-Aachen (blue),  Anti-kt (orange) and HDBSCAN (green). We use $R=0.3$ for Anti-kt and Cambridg-Aachen.}
    \label{fig:5}
\end{figure}

To validate the HDBSCAN clustering result, we compare the number of clustered subjets and the shape of subjets distributions obtained using Anti-kt and CA algorithms. Fig. \ref{fig:5} illustrates the number of clustered subjets by HDBSCAN alongside CA and Anti-kt for top jets (left plot) and QCD jets (right plot). HDBSCAN yields a larger number of clustered subjets, attributed to the dynamic nature of the jet radius. 
Furthermore, Figures \ref{fig:6} and \ref{fig:7} depict the kinematic distributions of the leading subjets for top and QCD jet events across all three algorithms. The energy distribution of the leading subjet distribution is softer for the HDBSCAN because HDBSCAN does not depend on the radius parameter and can identify the structure inside $R<0.3$ radius of CA  or anti-Kt subjets.

\section{Network performance}
\label{sec:5}
In this section, we compare the performance of the Mixer network against three baseline models: Particle Flow network (PFN) \cite{Komiske:2018cqr}, ParticleNet \cite{Qu:2019gqs}, and Particle Transformer network (ParT) \cite{Qu:2022mxj} for the top tagging task. 

PFN, rooted in Deep Sets, employs two symmetric neural networks to parameterize permutation-invariant symmetric functions. The first neural network comprises three fully connected (FC) layers with $100$, $100$, and $256$ neurons, respectively. The second neural network consists of three FC layers, each with $100$ neurons. 
The parameters of FC  layers are randomly initialized and activated by ReLU.  
The output layer, featuring two neurons, utilizes a softmax activation function.

The ParticleNet architecture integrates three Edge Convolution (EdgeConv) blocks. The initial EdgeConv block computes pairwise distances using the spatial coordinates of particles in the $(\eta-\phi)$ plane. Subsequent EdgeConv blocks employ the learned feature vectors as coordinates. 
The network updates particle information from its $16$ nearest neighbours. 
The dimensions of the EdgeCov blocks are $(64,64,64)$, $(128,128,128)$, and $(256,256,256)$, followed by channel-wise global average pooling to ensure the network's permutation invariance. A FC layer with $256$ neurons and ReLU activation processes the aggregated information from the EdgeCov blocks. Finally, a two-unit layer with softmax activation serves as the output layer.

ParT, based on attention-based transformer networks, comprises eight particle attention blocks and two class attention blocks with eight attention heads and a query dimension of $16$. Input feature interactions are encoded using four point-wise 1D convolutions with dimensions $(64,64,64,8)$. The attention blocks incorporate a $10\%$ dropout. A final output layer with two neurons and softmax activation completes the network.

For the Mixer network, we use the jet constituent dataset of the dimensions $(100, 7)$ and the subjet datasets with dimension $(15,7)$.  The mixer layer takes the jet constituent input and consists of one embedding layer with $128$ neurons. The first MLP consists of two FC layers with $128,64$ neurons and one Gaussian Error Linear Units (GELU) non-linear activation layer \cite{hendrycks2016gaussian}.
The second MLP consists of two FC layers with $64,128$ neurons and a GELU activation layer. The multi-heads cross-attention acts on the output of the second MLP, with dimension $(100,7)$, and the subjet dataset. The subjet dataset has $15$ parallel heads and a hidden dimension of size $64$. A dense layer with $64$ neurons is used with ReLU activation before the output layer.

\begin{figure*}[th!]
    \centering
    \includegraphics[scale=0.17]{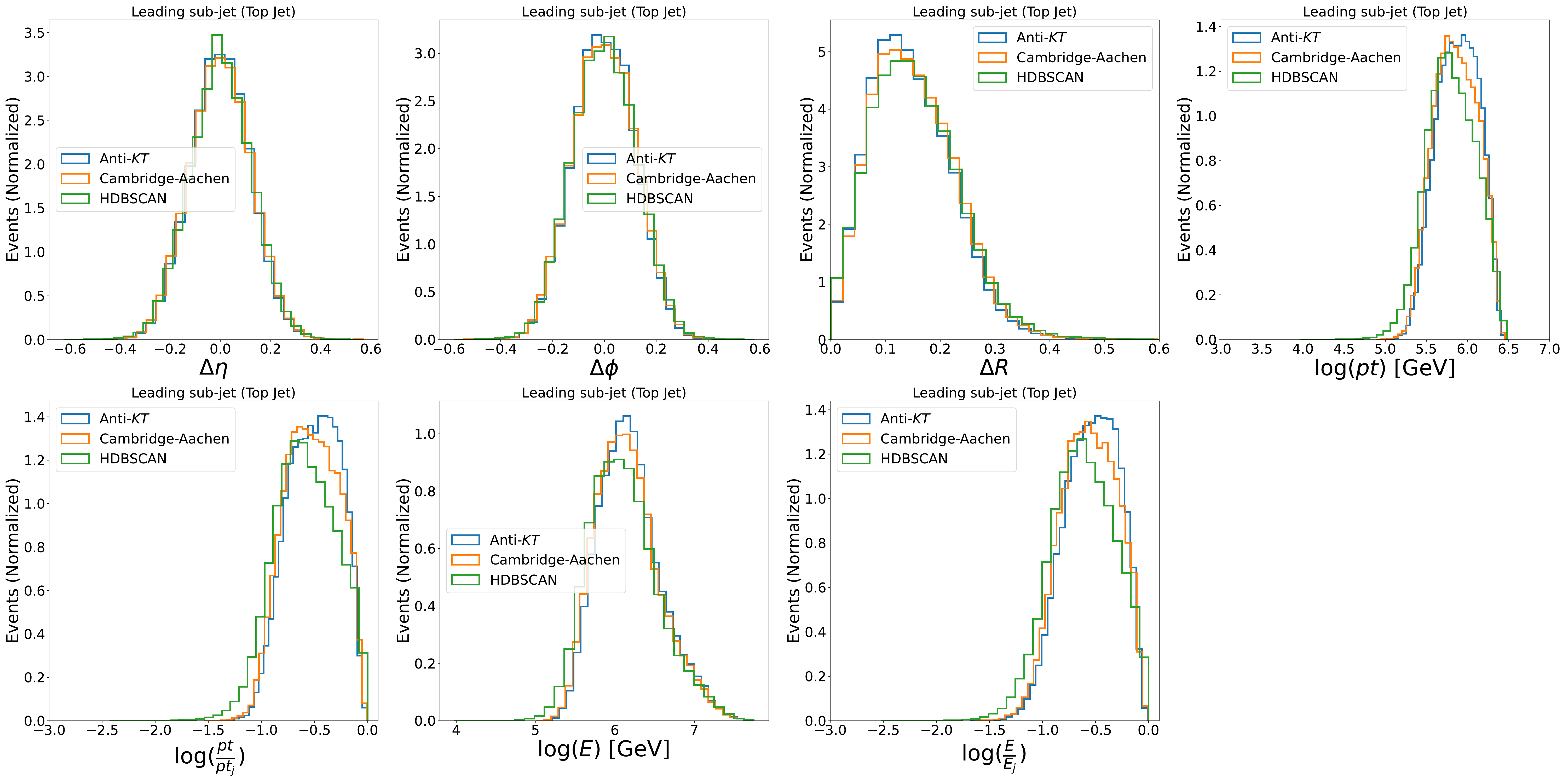}  
    \caption{Properties of the leading subjets of the top jet events clustered with Anti-kt (blue), CA (orange) and HDBSCAN (green). Anti-kt and CA are considered with radius parameter $R=0.3$.}
    \label{fig:6}
\end{figure*}

\begin{figure*}[th!]
    \centering
    \includegraphics[scale=0.17]{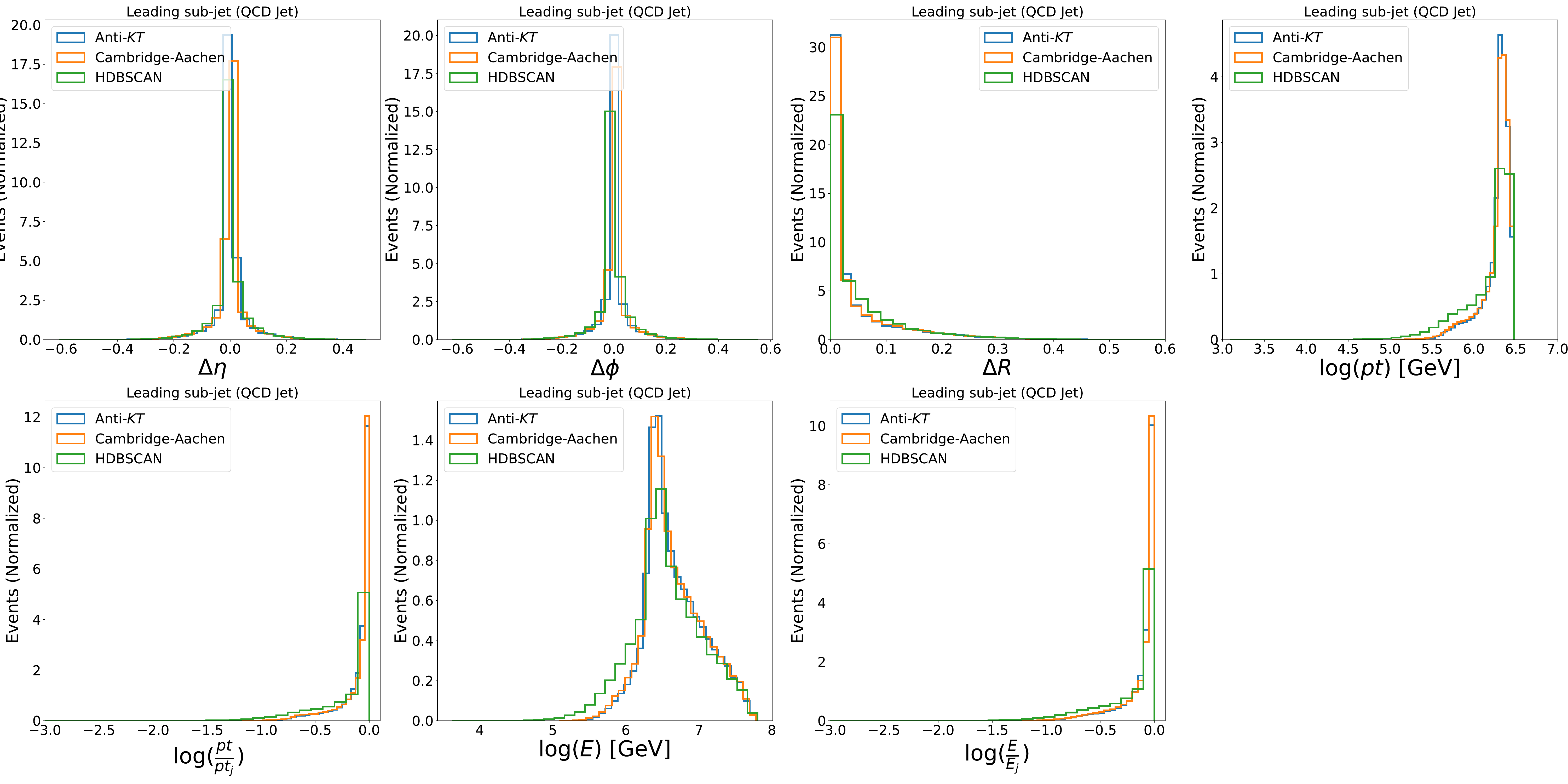}
    \caption{Properties of the leading subjets of the QCD jet events clustered with Anti-kt (blue), CA (orange) and HDBSCAN (green). For Anti-kt and CA, we use a radius parameter R=0.3.}
    \label{fig:7}
\end{figure*}
These networks are trained on the seven features extracted from the jet constituents of the top and QCD datasets, as detailed in Section \ref{sec:3}. Our network is trained on the same features but with an additional dataset encoding subjets information. This dataset is generated by applying jet clustering algorithms to cluster the jet constituents in the Top dataset, resulting in a secondary dataset comprising the seven features of all subjets. Figures \ref{fig:6} and \ref{fig:7} illustrate the considered features of the leading subjets for top and QCD jet events, respectively, when clustered using Anti-kt, CA, and HDBSCAN algorithms. 

To further test the impact of 
the mixer layer, we compare the network performance with a plain transformer model that analyzes the subjets dataset only. The architecture of the plain transformer is detailed in appendix \ref{app:A}.

\begin{table*}[htbp]
\centering
\caption{performance of the Mixer network for top quark tagging compared with other models. Results for EDI-net \cite{Moreno:2019bmu}, Point Cloud Transformer (PCT) \cite{Mikuni:2021pou}, Lorentz Net \cite{Gong:2022lye}, PELICAN \cite{Bogatskiy:2022czk}, PFN \cite{Komiske:2018cqr}, ParticleNET \cite{Qu:2019gqs}, and ParT \cite{Qu:2022mxj} are quoted from their published results. Pretrained ParticleNET and ParT have higher performance  with AUC = $0.9866$ and AUC= $0.9877$, respectively. The pertaining is done on the JETCLASS dataset, followed by the tuning to the top dataset. Transformer(subjet) model is trained from scratch using the CA subjets dataset only. Training time is per epoch with a batch size of $1024$.  The GPU training time is measured on an NVIDIA RTX A6000 card. 
}
\label{tab:1}
\begin{ruledtabular}
\begin{tabular}{cccccc}
                               &AUC & Rej$_{50\%}$ & Parameters & Time (GPU) [s] \\
   
    \hline
    JEDI-net with $\sum \mathcal{O}$ & $0.9807 $  &$-$& $87.7$K& $-$           \\
     \hline
    PFN                    & $0.9819$  &$247\pm 3$& $86.1$K& $\mathbf{30}$           \\
    \hline
     PCT         & $0.9855$  &$392\pm 7$& $193.3$K& $-$           \\
     \hline
     LorentzNet        & $0.9868$  &$498 \pm 18$& $224$K& $-$           \\
     \hline
    ParticleNET            & $0.9858$  &$397\pm 7$& $370$K& $-$             \\
    \hline
     PELICAN            & $\mathbf{0.9869}$  &$-$& $\mathbf{45}$K& $-$             \\
    \hline
    ParT                   & $0.9858$  & $413\pm 16$   &$2.14$M& $612$          \\
    \hline
    Transformer(subjets)   &  $0.9640$ & $186\pm 11$  & $398$K&  $129$   \\
    \hline
    Mixer(Anti-kt)         & $0.9854$  & $375\pm 5$  & $86.03$K& $33$          \\
    \hline
    Mixer(CA)              & $0.9856$  & $392\pm 6$  & $86.03$K& $33$           \\
    \hline
    Mixer(HDBSCAN)         & $0.9859$  & $416\pm 5$  & $86.03$K& $33$          \\
\end{tabular}
\end{ruledtabular}
\end{table*}

 Table \ref{tab:1} shows the network performances, the number of tunable parameters in each network, and the training time per epoch. Mixer network performances are reported according to the clustering algorithm of the subjet dataset. 
 For the subjet definition, HDBSCAN achieves the best performance over CA and Anti-kt as it can likely capture more geometrical information that improves network performance. 
  To achieve this performance, one needs both the mixer network for the jet constituent and the subjet inputs. This can be seen by comparing the performance of the transformer using subjet information alone, which shows the lowest classification performance. This underscores the necessity of incorporating MLPs in the mixer layer.
  The mixer networks are not only achieving state-of-the-art performance comparable to ParticleNET, PELICAN and ParT but also approximately 20 times faster in training.  PFN has the shortest training time but lacks learning of the local information shared between particles and their neighbours, leading to relatively poor performance.
\section{Interpretable ML techniques}
\label{sec:6}

ML models' interpretability can be challenging due to their intricate hidden layers. Understanding the model's architecture and learned representations is crucial for accurate predictions.

Various interpretable ML methods have been  developed to provide insights into how models make predictions. This helps to validate model decisions. 
In this section, we employ two methods
that offers a straightforward interpretation of the network outcomes, namely, Central Kernel Alignment (CKA) and attention map visualization. CKA  is a metric used to compare the similarity between two sets of learned representations in a high-dimensional feature space. It was first introduced in \cite{kornblith2019similarity} and used in collider analysis in \cite{Esmail:2023axd}. 

It measures the representations learned by the network layers or hidden layers of different models, considering local similarities and global structure.
On the other hand, attention maps are visual representations generated by attention mechanisms in neural networks, highlighting the input data most relevant for making predictions. They provide insights into the focuses of the model during processing, aiding in the interpretation of the decision-making process.

In the following, we apply those interpretable methods to the Mixer network trained on t a jet constituents dataset with dimensions $(100,7)$ and a subjets information with dimensions $(15,7)$  clustered using the CA algorithm with $R=0.3$. 
Importantly, these interpretable methods are agnostic to the specific network configuration and can be applied to other results presented in this paper.

\subsection{CKA similarity}
CKA similarity, rooted in the principles of kernel methods and alignment-based metrics, offers a comprehensive framework for assessing the similarity between two sets of representations learned by different models or layers within a model. It measures the alignment between representations in a high-dimensional feature space rather than simply comparing their values. 
Unlike linear similarity measures such as Pearson correlation or Euclidean distance, 

CKA captures complex relationships between representations learned by different models or layers, making it suitable for comparing high-dimensional and non-linearly transformed data. 
The primary obstacle in analyzing the representations of hidden layers in neural networks is the dispersion of features across neurons, with sizes often larger than the input dimension and varying in layers or models. 

CKA facilitates quantitative comparisons of representations both within individual networks and across different models. 
This can be done by considering the activation matrices of two hidden layers $X$ and $Y$ evaluated on the same input dataset; when the data size is $d$, and $P_1$ and $P_2$ is the number of neurons of the two different hidden layers, $X\in \mathbb{R}^{d\times P_1}$ and $Y\in \mathbb{R}^{d\times P_2}$. The CKA similarity is defined as 

\begin{equation}
    \text{CKA(M,N)} = \frac{\text{HSIC(M,N)}}{\sqrt{\text{HSIC(M,M)}\text{HSIC(N,N)}}}\,,
\end{equation}
where $M=XX^T$ and $N=YY^T$ are two Gram matrices of the two hidden layers with $d\times d$ dimension. 
The size of the Gram matrices depends only on the number of inputs, therefore, the $\text{CKA}(M,N)$ can be used 
to compare any layers with different numbers of neurons or networks of different models. 

The Hilbert-Schmidt Independent Criterion (HISC) \cite{greenfeld2020robust} between two matrices is defined as 
\begin{equation}
    \text{HSIC(M,N)} = \frac{1}{(d-1)^2} \text{Tr}(MHNH)\,,
\end{equation}
where  a $d\times d$ centering matrix $H$ is defined as $H_{ij}= \delta_{ij}-1/d$, therefore $\sum_i {AH}_{ij}=\sum_j{aH}_{ij}=0$ for $A=M, N$.  Centering the matrices 
ensures that the CKA similarity is not overly influenced by outliers or extreme values in the data, leading to more robust comparisons between representations. 

The value of the CKA ranges between $[0,1]$. 
A higher CKA value suggests that these layers have captured redundant information from the input features. If two subsequent layers are similar in the CKA, it indicates the second layer leads to negligible improvement in classification accuracy. In such instances, trimming these layers can reduce model complexity without compromising classification performance.
Conversely, the layers with lower CKA values have captured distinct information from the data, and enhanced the classification performance

\begin{figure}[th!]
    \centering
    \includegraphics[scale=0.2]{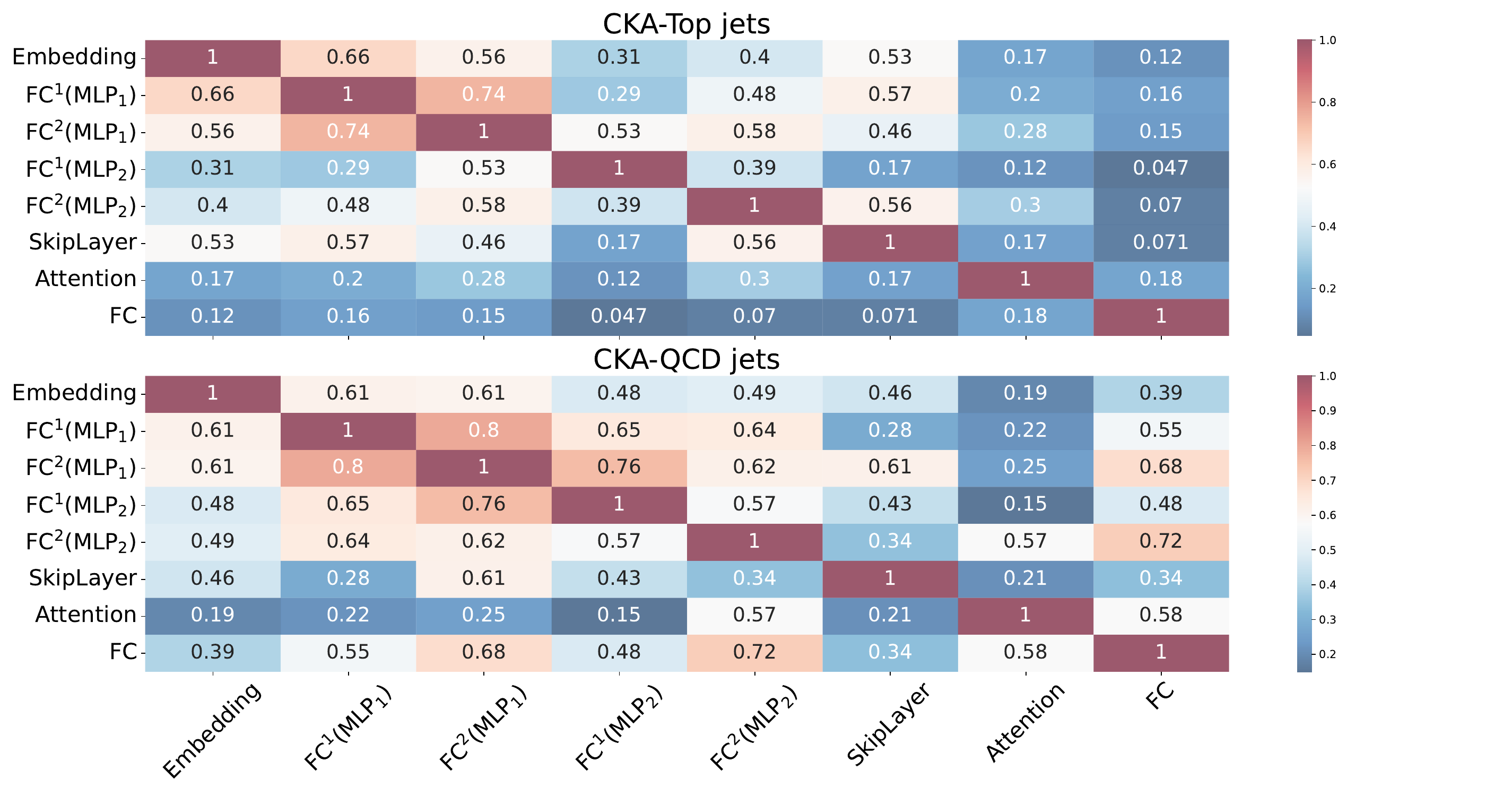}
    \caption{The CKA similarity of top jet events (top plot) and QCD jet (bottom). Axes represent the network layers.  FC(MLP$_1$) and FC(MLP$_2$) are the fully connected layers in the first and second MLP of mixer layers, respectively. The last FC represents the last FC layer in the network, and Attention is the multi-heads cross-attention.}
    \label{fig:8}
\end{figure}

The CKA results are depicted in Fig. \ref{fig:8}, showing the top jet events in the upper plot and QCD jet events in the lower plot. The analysis is based on a sample of $5000$ test events, with the subjets dataset clustered using the CA algorithm with $R=0.3$. CKA values are computed for distinctive model layers, including the embedding layer, the two FC layers for the first and second MLP mixer, the multi-head cross-attention layer, and the final FC layer. 

In general, layers with low correlations imply that they capture independent information from each other, underscoring their significance in the network's decision making process, see for example figure 3 in \cite{kornblith2019similarity}.

The multi-head cross-attention layer shows lower similarity with the two MLPs for the top jet with CKA value $30\%$ and $57\%$ for the QCD jets. The top jet CKA values are lower than QCD ones, which suggests the network layers are adept at capturing distinct information and are capable of learning the substructure of the top jet. The MLP mixer layers must have focused on the other features of the model.  The first and second MLP mixers exhibit low similarity.  Specifically, for top events, the two MLPs demonstrate lower CKA values around $58\%$ compared to the QCD events with CKA value $76\%$, suggesting that the network has learned a specific internal structure unique to top events.

\subsection{Attention maps}

Attention maps visualize the attention scores assigned
to each particle token in the input sequence, providing a representation of where the model focuses its attention during the decision making process\cite{chefer2021transformer}.  

Also, it reveals the relation between particle tokens. 

For instance, it highlights the information extracted from the jet constituents relevant to the clustered subjets. 
\begin{figure}[th!]
    \centering
    \includegraphics[scale=0.25]{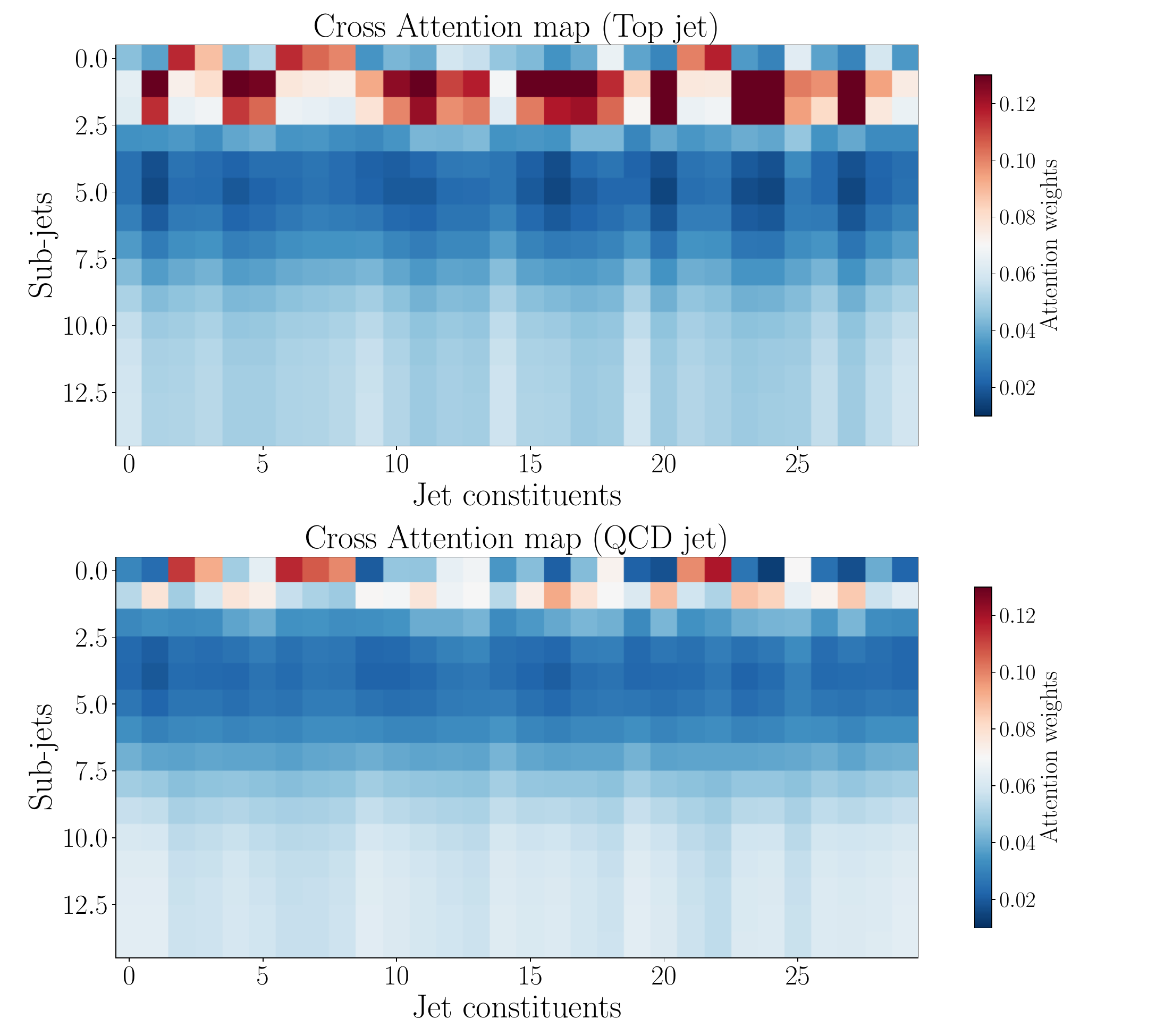}
    \caption{Cross-attention maps for 50000 test events of top (top plot) and QCD (bottom plot) 
    averaged over $15$ attention heads. The X-axis shows the attention score for the first transformed $30^\text{th}$ jet contents, while the Y-axis shows the
attention score for the transformed subjets.}
    \label{fig:9}
\end{figure}

Fig. \ref{fig:9} presents the cross-attention maps for a sample of 50,000 test events, showing top jet events in the upper plot and QCD jet events in the lower plot. As mentioned, the model is trained on two input datasets: jet constituents with dimensions $(100,7)$ and a subjets dataset with dimensions $(15,7)$ clustered by the CA algorithm with $R=0.3$. The network comprises 15 cross-attention heads that operate in parallel, with Fig. \ref{fig:9} displaying the average output of these 15 heads. The cross-attention maps for all attention heads individually are illustrated in Appendix \ref{app:B}. 

The visualizations reveal a distinct focus of the network; it concentrates on three subjets to identify top events, while it directs its attention to only one subjet for identifying QCD events. This observation underscores the effectiveness of the multi-head cross-attention mechanism, particularly in conjunction with the subjets dataset, for capturing the substructure inherent in top jets. In the appendix, we also see the 4th-8th subjets contribute to individual attention heads, consistent with the previous ML study based on the substructure variables. 

\section{\bf {Conclusion}}
In this paper, we present a simple permutation-invariant network, a "Mixer network" assisted by subjet information. 
The "Mixer network" has a simple structure with fewer tunable parameters and is approximately 20 times faster than state-of-the-art networks, such as ParticleNET and  ParT,  with comparable classification performance.

The network comprises mixer layers designed to maintain the dimensions of both input and output datasets. This allows for easy scalability of model complexity by stacking additional mixer layers, thereby enhancing expressivity when analyzing intricate data. Each mixer layer consists of two MLPs, with weight sharing across the entire dataset, facilitating the mixing of particle and feature tokens within the particle cloud. As a result, these MLPs can capture global and local features, including relationships between individual particles and their nearest neighbours. Subjet information is utilized to improve further the network's ability to learn the local structure of the jet. Cross-attention mechanisms are employed to analyze both the subjets and the jet constituents datasets, enabling the extraction of relevant local information and updating model weights to ensure that each jet constituent incorporates updated information about its nearby particles.

The secondary dataset is generated using three jet clustering algorithms: Anti-kt, CA, and HDBSCAN. HDBSCAN is a density-based clustering algorithm which does not depend on any radius parameters. This paper introduces HDBSCAN in collider analysis for the first time and presents a detailed comparison with recursive clustering algorithms such as Anti-kt and CA. 
Employing the Mixer network with a second dataset clustered by HDBSCAN demonstrates relatively higher classification performance than anti-kt or CA.

To elucidate the network results,  CKA similarity and attention map visualization are utilized. CKA analysis confirms that the cross-attention and mixer layer capture the different information. Moreover, the first MLP layers, responsible for mixing particle tokens, and the second MLP layers, responsible for mixing feature tokens, learn distinct information, thereby ensuring their effectiveness in capturing the substructure of the jet.
Moreover, visualization of the cross-attention maps reveals that the network assigns higher weights to the leading subjets and their associated jet constituents, underscoring the efficacy of cross-attention mechanisms in incorporating neighbouring information into each jet constituent. 

The merit of using cross-attention arises from the fact that the jet substructure is conditioned by a few partons originating from the hard process; therefore, the estimated probability of event distribution can be expressed by the product of the hard parton distribution ($\sim$ jets or subjet distribution) times the hadron distribution in the jet conditioned by the parton distribution. 
Cross-attention guarantees that subjet and jet constituent information appear as multiplicative matrices.

In \cite{Hammad:2023sbd}, we used cross-attention heads to connect the outputs of transformer layers of global event kinematics and that of jet substructures. 
It has shown that the network effectively learned the correlation between jet substructures and event kinematics, significantly improving the event classification task of $pp\rightarrow H \rightarrow hh$. 
One of the drawbacks of this approach comes from the computational complexity of using transformer layers for the jet encoding step. In this paper, we drastically reduced the network complexity by replacing the transformer layer with the subjet-assisted MLP mixer. 
The output of the mixer layer has the same dimensions as the input dataset. Therefore, we can replace the transformer layer of subjet analysis in  \cite{Hammad:2023sbd} with the mixer layer, keeping the same performance.   In short, the network proposed in this paper opens up the step toward global event analysis encoded in the particle cloud much more efficiently than the previous state-of-the-art approaches.


\section*{Acknowledgments}
 This work is funded by grant number 22H05113, ``Foundation of Machine Learning Physics'', Grant in Aid for Transformative Research Areas and 22K03626, Grant-in-Aid for Scientific Research (C).
\appendix
\section*{A. Plain transformer architecture}
\label{app:A}
We employ a straightforward attention-based transformer model with multi-head self-attention to analyze the subjet dataset clustered using the CA algorithm. The model takes as input a subjet dataset with dimensions of $(15,7)$, where $15$ represents the number of subjet tokens and $7$ denotes the corresponding features.

Comprising three stacked transformer layers and an MLP with three FC layers, each transformer layer features $8$ self-attention heads operating in parallel, with a hidden dimension,  query dimension, set to $256$. 
The output from these attention heads is then combined with the original input data via a skip connection layer. 
The resulting output from the skip connection undergoes flattening and is passed through two fully connected layers with $128$ and $7$ neurons, respectively, utilizing the GELU activation function. 
Subsequently, the output from the final fully connected layer is combined with the self-attention output via a second skip connection layer.

Following this, the final output of the transformer layer undergoes normalization and retains the same dimension as the input dataset. The output of the transformer layers is then fed through an MLP comprising three fully connected layers with dimensions of $256$, $128$, and $64$, employing the GELU activation function. 
After each fully connected layer, a dropout layer with a dropout rate of $20\%$ is applied to prevent overfitting. 
Finally, the output is passed to the output layer with two neurons and softmax activation for classification.
The model is trained for $30$ epochs with a batch size of $1024$.

\section*{B. Attention heads visualization}
\label{app:B}
\begin{figure*}[htbp]
    \centering
    \includegraphics[scale=0.34]{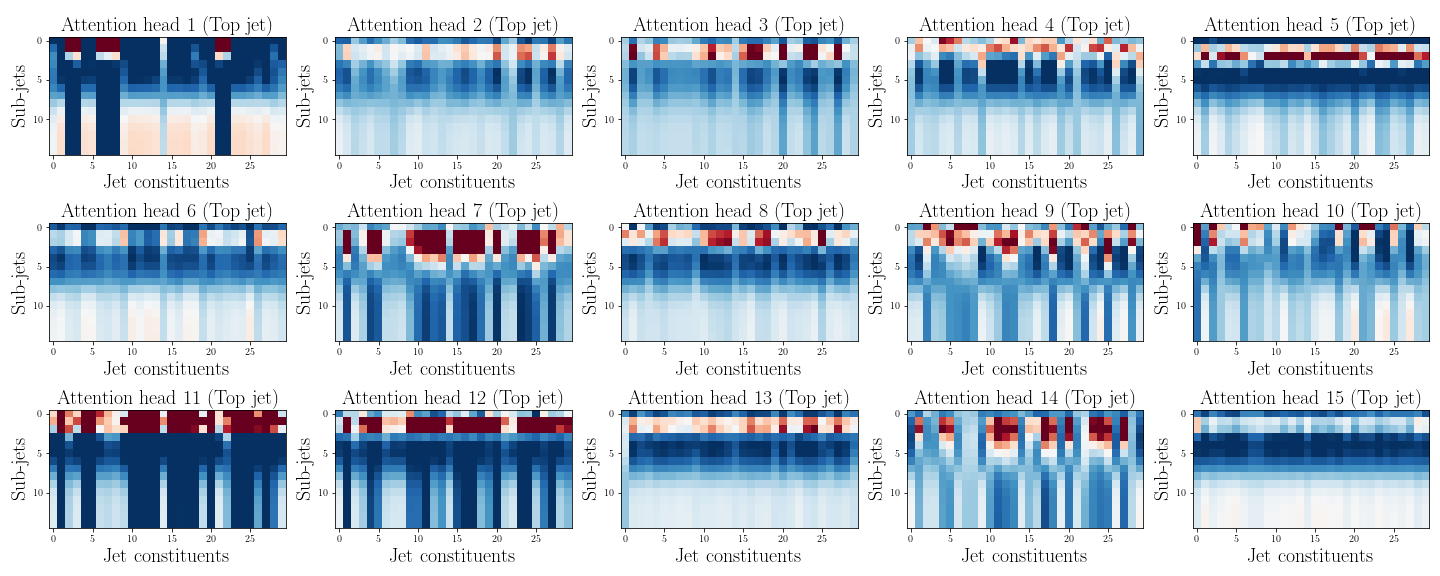} \\
    \includegraphics[scale=0.34]{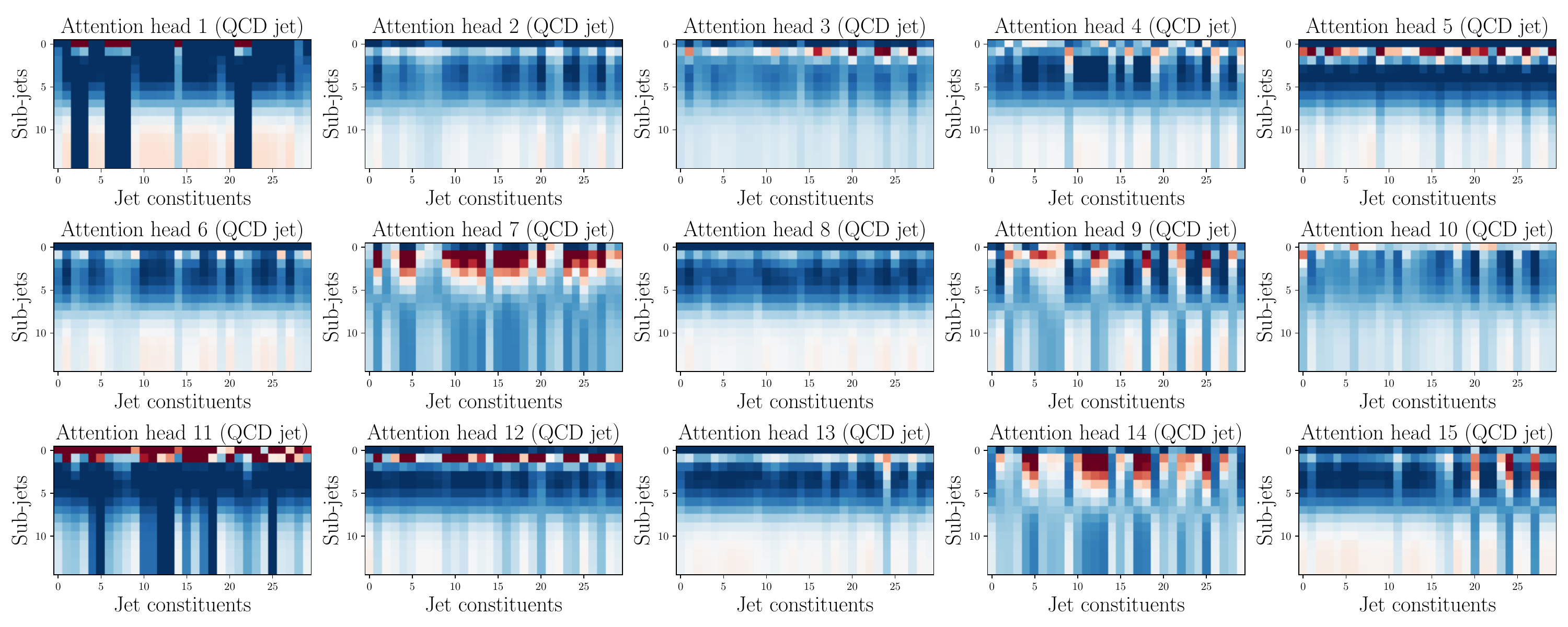}
    \caption{Cross-attention maps of the $15$ attention heads for $50000$ test events. The upper three rows are for top jet events, and the bottom rows are fro QCD events. }
    \label{fig:10}
\end{figure*}

In this appendix, we present the cross-attention maps of the 15 utilized attention heads individually. The output of these cross-attention heads possesses dimensions of $(N, 15, 80, 15)$, where $N$ denotes the number of test events, 15 signifies the number of parallel attention heads, and the last two dimensions indicate the quantities of jet constituents and subjet tokens, respectively.

In Fig. \ref{fig:10}, we depict the accumulated averages of 50,000 test events for each attention head separately. For clarity, we maintain the first 30 jet constituents along the $X$ axis, with the $Y$ axis representing the subjet tokens. Notably, Fig. \ref{fig:10} reveals that the Mixer network assigns considerable attention to a larger number of subleading subjets. The detail was not apparent in Fig. \ref{fig:9}, where we averaged across all heads.
\bibliographystyle{unsrt}
\bibliography{biblo}

\begin{thebibliography}{10}

\bibitem{Butterworth:2008iy}
Jonathan~M. Butterworth, Adam~R. Davison, Mathieu Rubin, and Gavin~P. Salam.
\newblock {Jet substructure as a new Higgs search channel at the LHC}.
\newblock {\em Phys. Rev. Lett.}, 100:242001, 2008.

\bibitem{Kaplan:2008ie}
David~E. Kaplan, Keith Rehermann, Matthew~D. Schwartz, and Brock Tweedie.
\newblock {Top Tagging: A Method for Identifying Boosted Hadronically Decaying
  Top Quarks}.
\newblock {\em Phys. Rev. Lett.}, 101:142001, 2008.

\bibitem{Cui:2010km}
Yanou Cui, Zhenyu Han, and Matthew~D. Schwartz.
\newblock {W-jet Tagging: Optimizing the Identification of Boosted
  Hadronically-Decaying W Bosons}.
\newblock {\em Phys. Rev. D}, 83:074023, 2011.

\bibitem{Plehn:2011sj}
Tilman Plehn, Michael Spannowsky, and Michihisa Takeuchi.
\newblock {How to Improve Top Tagging}.
\newblock {\em Phys. Rev. D}, 85:034029, 2012.

\bibitem{Soper:2012pb}
Davison~E. Soper and Michael Spannowsky.
\newblock {Finding top quarks with shower deconstruction}.
\newblock {\em Phys. Rev. D}, 87:054012, 2013.

\bibitem{Anders:2013oga}
Christoph Anders, Catherine Bernaciak, Gregor Kasieczka, Tilman Plehn, and
  Torben Schell.
\newblock {Benchmarking an even better top tagger algorithm}.
\newblock {\em Phys. Rev. D}, 89(7):074047, 2014.

\bibitem{Kasieczka:2015jma}
Gregor Kasieczka, Tilman Plehn, Torben Schell, Thomas Strebler, and Gavin~P.
  Salam.
\newblock {Resonance Searches with an Updated Top Tagger}.
\newblock {\em JHEP}, 06:203, 2015.

\bibitem{Thaler:2010tr}
Jesse Thaler and Ken Van~Tilburg.
\newblock {Identifying Boosted Objects with N-subjettiness}.
\newblock {\em JHEP}, 03:015, 2011.

\bibitem{Thaler:2011gf}
Jesse Thaler and Ken Van~Tilburg.
\newblock {Maximizing Boosted Top Identification by Minimizing N-subjettiness}.
\newblock {\em JHEP}, 02:093, 2012.

\bibitem{Larkoski:2013eya}
Andrew~J. Larkoski, Gavin~P. Salam, and Jesse Thaler.
\newblock {Energy Correlation Functions for Jet Substructure}.
\newblock {\em JHEP}, 06:108, 2013.

\bibitem{Moult:2016cvt}
Ian Moult, Lina Necib, and Jesse Thaler.
\newblock {New Angles on Energy Correlation Functions}.
\newblock {\em JHEP}, 12:153, 2016.

\bibitem{Larkoski:2014wba}
Andrew~J. Larkoski, Simone Marzani, Gregory Soyez, and Jesse Thaler.
\newblock {Soft Drop}.
\newblock {\em JHEP}, 05:146, 2014.

\bibitem{Abdesselam:2010pt}
A.~Abdesselam et~al.
\newblock {Boosted Objects: A Probe of Beyond the Standard Model Physics}.
\newblock {\em Eur. Phys. J. C}, 71:1661, 2011.

\bibitem{Altheimer:2012mn}
A.~Altheimer et~al.
\newblock {Jet Substructure at the Tevatron and LHC: New results, new tools,
  new benchmarks}.
\newblock {\em J. Phys. G}, 39:063001, 2012.

\bibitem{Altheimer:2013yza}
A.~Altheimer et~al.
\newblock {Boosted Objects and Jet Substructure at the LHC. Report of
  BOOST2012, held at IFIC Valencia, 23rd-27th of July 2012}.
\newblock {\em Eur. Phys. J. C}, 74(3):2792, 2014.

\bibitem{Cogan:2014oua}
Josh Cogan, Michael Kagan, Emanuel Strauss, and Ariel Schwarztman.
\newblock {Jet-Images: Computer Vision Inspired Techniques for Jet Tagging}.
\newblock {\em JHEP}, 02:118, 2015.

\bibitem{Almeida:2015jua}
Leandro~G. Almeida, Mihailo Backovi\'c, Mathieu Cliche, Seung~J. Lee, and Maxim
  Perelstein.
\newblock {Playing Tag with ANN: Boosted Top Identification with Pattern
  Recognition}.
\newblock {\em JHEP}, 07:086, 2015.

\bibitem{deOliveira:2015xxd}
Luke de~Oliveira, Michael Kagan, Lester Mackey, Benjamin Nachman, and Ariel
  Schwartzman.
\newblock {Jet-images \textemdash{} deep learning edition}.
\newblock {\em JHEP}, 07:069, 2016.

\bibitem{Baldi:2016fql}
Pierre Baldi, Kevin Bauer, Clara Eng, Peter Sadowski, and Daniel Whiteson.
\newblock {Jet Substructure Classification in High-Energy Physics with Deep
  Neural Networks}.
\newblock {\em Phys. Rev. D}, 93(9):094034, 2016.

\bibitem{Barnard:2016qma}
James Barnard, Edmund~Noel Dawe, Matthew~J. Dolan, and Nina Rajcic.
\newblock {Parton Shower Uncertainties in Jet Substructure Analyses with Deep
  Neural Networks}.
\newblock {\em Phys. Rev. D}, 95(1):014018, 2017.

\bibitem{Komiske:2016rsd}
Patrick~T. Komiske, Eric~M. Metodiev, and Matthew~D. Schwartz.
\newblock {Deep learning in color: towards automated quark/gluon jet
  discrimination}.
\newblock {\em JHEP}, 01:110, 2017.

\bibitem{Kasieczka:2017nvn}
Gregor Kasieczka, Tilman Plehn, Michael Russell, and Torben Schell.
\newblock {Deep-learning Top Taggers or The End of QCD?}
\newblock {\em JHEP}, 05:006, 2017.

\bibitem{Macaluso:2018tck}
Sebastian Macaluso and David Shih.
\newblock {Pulling Out All the Tops with Computer Vision and Deep Learning}.
\newblock {\em JHEP}, 10:121, 2018.

\bibitem{Choi:2018dag}
Suyong Choi, Seung~J. Lee, and Maxim Perelstein.
\newblock {Infrared Safety of a Neural-Net Top Tagging Algorithm}.
\newblock {\em JHEP}, 02:132, 2019.

\bibitem{Shlomi:2020gdn}
Jonathan Shlomi, Peter Battaglia, and Jean-Roch Vlimant.
\newblock {Graph Neural Networks in Particle Physics}.
\newblock {\em Mach. Learn.: Sci. Technol}, 2:021001, 7 2021.

\bibitem{Mokhtar:2022pwm}
Farouk Mokhtar, Raghav Kansal, and Javier Duarte.
\newblock {Do graph neural networks learn traditional jet substructure?}
\newblock In {\em {36th Conference on Neural Information Processing Systems}:
  {Workshop on Machine Learning and the Physical Sciences}}, 11 2022.

\bibitem{Ma:2022bvt}
Fei Ma, Feiyi Liu, and Wei Li.
\newblock {Jet tagging algorithm of graph network with Haar pooling message
  passing}.
\newblock {\em Phys. Rev. D}, 108(7):072007, 2023.

\bibitem{Gong:2022lye}
Shiqi Gong, Qi~Meng, Jue Zhang, Huilin Qu, Congqiao Li, Sitian Qian, Weitao Du,
  Zhi-Ming Ma, and Tie-Yan Liu.
\newblock {An efficient Lorentz equivariant graph neural network for jet
  tagging}.
\newblock {\em JHEP}, 07:030, 2022.

\bibitem{Dreyer:2020brq}
Fr\'ed\'eric~A. Dreyer and Huilin Qu.
\newblock {Jet tagging in the Lund plane with graph networks}.
\newblock {\em JHEP}, 03:052, 2021.

\bibitem{Guest:2016iqz}
Daniel Guest, Julian Collado, Pierre Baldi, Shih-Chieh Hsu, Gregor Urban, and
  Daniel Whiteson.
\newblock {Jet Flavor Classification in High-Energy Physics with Deep Neural
  Networks}.
\newblock {\em Phys. Rev. D}, 94(11):112002, 2016.

\bibitem{Pearkes:2017hku}
Jannicke Pearkes, Wojciech Fedorko, Alison Lister, and Colin Gay.
\newblock {Jet Constituents for Deep Neural Network Based Top Quark Tagging}.
\newblock 4 2017.

\bibitem{Egan:2017ojy}
Shannon Egan, Wojciech Fedorko, Alison Lister, Jannicke Pearkes, and Colin Gay.
\newblock {Long Short-Term Memory (LSTM) networks with jet constituents for
  boosted top tagging at the LHC}.
\newblock 11 2017.

\bibitem{Fraser:2018ieu}
Katherine Fraser and Matthew~D. Schwartz.
\newblock {Jet Charge and Machine Learning}.
\newblock {\em JHEP}, 10:093, 2018.

\bibitem{Butter:2017cot}
Anja Butter, Gregor Kasieczka, Tilman Plehn, and Michael Russell.
\newblock {Deep-learned Top Tagging with a Lorentz Layer}.
\newblock {\em SciPost Phys.}, 5(3):028, 2018.

\bibitem{Kasieczka:2018lwf}
Gregor Kasieczka, Nicholas Kiefer, Tilman Plehn, and Jennifer~M. Thompson.
\newblock {Quark-Gluon Tagging: Machine Learning vs Detector}.
\newblock {\em SciPost Phys.}, 6(6):069, 2019.

\bibitem{Komiske:2018cqr}
Patrick~T. Komiske, Eric~M. Metodiev, and Jesse Thaler.
\newblock {Energy Flow Networks: Deep Sets for Particle Jets}.
\newblock {\em JHEP}, 01:121, 2019.

\bibitem{Qu:2019gqs}
Huilin Qu and Loukas Gouskos.
\newblock {ParticleNet: Jet Tagging via Particle Clouds}.
\newblock {\em Phys. Rev. D}, 101(5):056019, 2020.

\bibitem{Qu:2022mxj}
Huilin Qu, Congqiao Li, and Sitian Qian.
\newblock {Particle Transformer for Jet Tagging}.
\newblock {\em Proceedings of the 39th International Conference on Machine
  Learning(PMLR)}, 162:18281--18292, 2022.

\bibitem{Finke:2023veq}
Thorben Finke, Michael Kr\"amer, Alexander M\"uck, and Jan T\"onshoff.
\newblock {Learning the language of QCD jets with transformers}.
\newblock {\em JHEP}, 06:184, 2023.

\bibitem{Shmakov:2021qdz}
Alexander Shmakov, Michael~James Fenton, Ta-Wei Ho, Shih-Chieh Hsu, Daniel
  Whiteson, and Pierre Baldi.
\newblock {SPANet: Generalized permutationless set assignment for particle
  physics using symmetry preserving attention}.
\newblock {\em SciPost Phys.}, 12(5):178, 2022.

\bibitem{Hammad:2023sbd}
A.~Hammad, S.~Moretti, and M.~Nojiri.
\newblock {Multi-scale cross-attention transformer encoder for event
  classification}.
\newblock {\em JHEP}, 03:144, 2024.

\bibitem{He:2023cfc}
Minxuan He and Daohan Wang.
\newblock {Quark/gluon discrimination and top tagging with dual attention
  transformer}.
\newblock {\em Eur. Phys. J. C}, 83(12):1116, 2023.

\bibitem{zaheer2017deep}
Manzil Zaheer, Satwik Kottur, Siamak Ravanbakhsh, Barnabas Poczos, Russ~R
  Salakhutdinov, and Alexander~J Smola.
\newblock Deep sets.
\newblock {\em Advances in neural information processing systems}, 30, 2017.

\bibitem{Moreno:2019bmu}
Eric~A. Moreno, Olmo Cerri, Javier~M. Duarte, Harvey~B. Newman, Thong~Q.
  Nguyen, Avikar Periwal, Maurizio Pierini, Aidana Serikova, Maria Spiropulu,
  and Jean-Roch Vlimant.
\newblock {JEDI-net: a jet identification algorithm based on interaction
  networks}.
\newblock {\em Eur. Phys. J. C}, 80(1):58, 2020.

\bibitem{Mikuni:2021pou}
Vinicius Mikuni and Florencia Canelli.
\newblock {Point cloud transformers applied to collider physics}.
\newblock {\em Mach. Learn. Sci. Tech.}, 2(3):035027, 2021.

\bibitem{Bogatskiy:2022czk}
Alexander Bogatskiy, Timothy Hoffman, David~W. Miller, and Jan~T. Offermann.
\newblock {PELICAN: Permutation Equivariant and Lorentz Invariant or Covariant
  Aggregator Network for Particle Physics}.
\newblock 11 2022.

\bibitem{vaswani2017attention}
Ashish Vaswani, Noam Shazeer, Niki Parmar, Jakob Uszkoreit, Llion Jones,
  Aidan~N Gomez, {\L}ukasz Kaiser, and Illia Polosukhin.
\newblock Attention is all you need.
\newblock {\em Advances in neural information processing systems}, 30, 2017.

\bibitem{tolstikhin2021mlp}
Ilya~O Tolstikhin, Neil Houlsby, Alexander Kolesnikov, Lucas Beyer, Xiaohua
  Zhai, Thomas Unterthiner, Jessica Yung, Andreas Steiner, Daniel Keysers,
  Jakob Uszkoreit, et~al.
\newblock Mlp-mixer: An all-mlp architecture for vision.
\newblock {\em Advances in neural information processing systems},
  34:24261--24272, 2021.

\bibitem{Buhmann:2023acn}
Erik Buhmann, Cedric Ewen, Gregor Kasieczka, Vinicius Mikuni, Benjamin Nachman,
  and David Shih.
\newblock {Full phase space resonant anomaly detection}.
\newblock {\em Phys. Rev. D}, 109(5):055015, 2024.

\bibitem{Dokshitzer:1997in}
Yuri~L. Dokshitzer, G.~D. Leder, S.~Moretti, and B.~R. Webber.
\newblock {Better jet clustering algorithms}.
\newblock {\em JHEP}, 08:001, 1997.

\bibitem{Cacciari:2008gp}
Matteo Cacciari, Gavin~P. Salam, and Gregory Soyez.
\newblock {The anti-$k_t$ jet clustering algorithm}.
\newblock {\em JHEP}, 04:063, 2008.

\bibitem{10.1007/978-3-642-37456-2_14}
Ricardo J. G.~B. Campello, Davoud Moulavi, and Joerg Sander.
\newblock Density-based clustering based on hierarchical density estimates.
\newblock In Jian Pei, Vincent~S. Tseng, Longbing Cao, Hiroshi Motoda, and
  Guandong Xu, editors, {\em Advances in Knowledge Discovery and Data Mining},
  pages 160--172, Berlin, Heidelberg, 2013. Springer Berlin Heidelberg.

\bibitem{qi2017pointnet}
Charles~R Qi, Hao Su, Kaichun Mo, and Leonidas~J Guibas.
\newblock Pointnet: Deep learning on point sets for 3d classification and
  segmentation.
\newblock In {\em Proceedings of the IEEE conference on computer vision and
  pattern recognition}, pages 652--660, 2017.

\bibitem{Walsh:2011fz}
Jonathan~R. Walsh and Saba Zuberi.
\newblock {Factorization Constraints on Jet Substructure}.
\newblock 10 2011.

\bibitem{Bierlich:2022pfr}
Christian Bierlich et~al.
\newblock {A comprehensive guide to the physics and usage of PYTHIA 8.3}.
\newblock {\em SciPost Phys. Codeb.}, 2022:8, 2022.

\bibitem{deFavereau:2013fsa}
J.~de~Favereau, C.~Delaere, P.~Demin, A.~Giammanco, V.~Lema\^\i{}tre,
  A.~Mertens, and M.~Selvaggi.
\newblock {DELPHES 3, A modular framework for fast simulation of a generic
  collider experiment}.
\newblock {\em JHEP}, 02:057, 2014.

\bibitem{Li:2022xfc}
Congqiao Li, Huilin Qu, Sitian Qian, Qi~Meng, Shiqi Gong, Jue Zhang, Tie-Yan
  Liu, and Qiang Li.
\newblock {Does Lorentz-symmetric design boost network performance in jet
  physics?}
\newblock {\em Phys. Rev. D}, 109(5):056003, 2024.

\bibitem{Cacciari:2011ma}
Matteo Cacciari, Gavin~P. Salam, and Gregory Soyez.
\newblock {FastJet User Manual}.
\newblock {\em Eur. Phys. J. C}, 72:1896, 2012.

\bibitem{Cerro:2021abp}
Giorgio Cerro, Srinandan Dasmahapatra, Henry~A. Day-Hall, Billy Ford, Stefano
  Moretti, and Claire~H. Shepherd-Themistocleous.
\newblock {Spectral clustering for jet physics}.
\newblock {\em JHEP}, 02:165, 2022.

\bibitem{Mukhopadhyaya:2023rsb}
Biswarup Mukhopadhyaya, Tousik Samui, and Ritesh~K. Singh.
\newblock {Dynamic radius jet clustering algorithm}.
\newblock {\em JHEP}, 04:019, 2023.

\bibitem{Komiske:2017ubm}
Patrick~T. Komiske, Eric~M. Metodiev, Benjamin Nachman, and Matthew~D.
  Schwartz.
\newblock {Pileup Mitigation with Machine Learning (PUMML)}.
\newblock {\em JHEP}, 12:051, 2017.

\bibitem{hendrycks2016gaussian}
Dan Hendrycks and Kevin Gimpel.
\newblock Gaussian error linear units (gelus).
\newblock {\em arXiv preprint arXiv:1606.08415}, 2016.

\bibitem{kornblith2019similarity}
Simon Kornblith, Mohammad Norouzi, Honglak Lee, and Geoffrey Hinton.
\newblock Similarity of neural network representations revisited.
\newblock In {\em International conference on machine learning}, pages
  3519--3529. PMLR, 2019.

\bibitem{Esmail:2023axd}
W.~Esmail, A.~Hammad, and S.~Moretti.
\newblock {Sharpening the A \textrightarrow{} Z$^{(*)}$h signature of the
  Type-II 2HDM at the LHC through advanced Machine Learning}.
\newblock {\em JHEP}, 11:020, 2023.

\bibitem{greenfeld2020robust}
Daniel Greenfeld and Uri Shalit.
\newblock Robust learning with the hilbert-schmidt independence criterion.
\newblock In {\em International Conference on Machine Learning}, pages
  3759--3768. PMLR, 2020.

\bibitem{chefer2021transformer}
Hila Chefer, Shir Gur, and Lior Wolf.
\newblock Transformer interpretability beyond attention visualization.
\newblock In {\em Proceedings of the IEEE/CVF conference on computer vision and
  pattern recognition}, pages 782--791, 2021.

\end{thebibliography}
\end{document}